\documentclass[11pt,a4paper]{article}
\pdfoutput=1
\usepackage{a4wide}
\usepackage{amsmath}
\usepackage{amssymb}
\usepackage{amsfonts}
\usepackage{cite}
\usepackage[table]{xcolor}
\usepackage{adjustbox}
\usepackage{multirow}
\usepackage{pdflscape}
\usepackage{gensymb}
\usepackage{graphicx}
\usepackage{diagbox}
\usepackage{pifont}
\usepackage{bigstrut}
\usepackage[small,bf]{caption}
\usepackage{enumerate}
\usepackage{tabulary}
\usepackage{float}
\usepackage{arydshln}
\usepackage{enumitem}
\usepackage{mathtools}
\usepackage[utf8]{inputenc}
\usepackage[T1]{fontenc}
\usepackage{footnote}
\usepackage{footmisc}
\usepackage{hyperref}
\usepackage{ctable}
\newcommand{\dcp}{\delta_{\mathrm{CP}}}

\def\1{\mathbf{1}}
\def\3{\mathbf{3}}
\def\2{\mathbf{2}}

\numberwithin{equation}{section}
\makeatletter
\g@addto@macro\bfseries{\boldmath}
\makeatother

\newcounter{savefootnote}
\newcounter{symfootnote}
\newcommand{\symfootnote}[1]{%
   \setcounter{savefootnote}{\value{footnote}}%
   \setcounter{footnote}{\value{symfootnote}}%
   \ifnum\value{footnote}>8\setcounter{footnote}{0}\fi%
   \let\oldthefootnote=\thefootnote%
   \renewcommand{\thefootnote}{\fnsymbol{footnote}}%
   \footnote{#1}%
   \let\thefootnote=\oldthefootnote%
   \setcounter{symfootnote}{\value{footnote}}%
   \setcounter{footnote}{\value{savefootnote}}%
}

\begin{document}
%=============

\begin{titlepage}

\vspace*{-15mm}

\begin{center}

{\bf\LARGE{Effect of Matter Density in T2HK and DUNE}}\\[10mm]

Monojit Ghosh,$^{a,b}$\footnote{E-mail: \texttt{mghosh@irb.hr}} 
and
Osamu Yasuda$^{\,c,}$\footnote{E-mail: \texttt{yasuda@phys.se.tmu.ac.jp}} \\
\vspace{8mm}
$^{a}$\,{\it School of Physics, University of Hyderabad, Hyderabad - 500046, India} \\
\vspace{2mm}
$^{b}$\,{\it Center of Excellence for Advanced Materials and Sensing Devices, Ruder Bo\v{s}kovi\'c Institute, 10000 Zagreb, Croatia} \\
\vspace{2mm}
$^{c}$\,{\it Department of Physics, Tokyo Metropolitan University, Hachioji, Tokyo 192-0397, Japan} \\

\end{center}
\vspace{8mm}

\begin{abstract}
\noindent 
CP phase determination for the near future long baseline experiments, T2HK and DUNE, will require precise measurements of the oscillation probabilities. However, the uncertainty in the Earth's density must be considered in determining these oscillation probabilities.  Therefore, in this study, we update the individual sensitivities of these experiments for determining the current unknowns in the standard three flavor scenario considering the latest configuration and also the complementarity between them while considering the uncertainty in the density. Our study showed that this uncertainty has a non-negligible impact on the precision of the CP phase determination particularly for DUNE.

\end{abstract}
\end{titlepage}

\setcounter{footnote}{0}

\section{Introduction}

Many neutrino experiments over the last twenty years have successfully determined the properties of neutrino -- such as mixing angles and mass squared differences --\!\cite{Workman:2022ynf}. However, the quantities yet to be determined include the mass ordering of the three neutrinos, the octant to which $\theta_{23}$ belongs, and the leptonic CP phase. These unknown quantities are expected to be determined by the near future long baseline experiments, which are T2HK~\cite{Hyper-Kamiokande:2016srs} and DUNE~\cite{DUNE:2021cuw}. The standard representation of leptonic mixing is given by a $3\times 3$ unitary matrix
\begin{eqnarray}
&{\ }& \hspace{-10mm}
  U\equiv
\left(
\begin{array}{ccc}
 1 &0 &0 \\
0  &c_{23} &s_{23} \\
0  &-s_{23}  &c_{23}  \\
\end{array}
\right)
\left(
\begin{array}{ccc}
c_{13}  &0 &s_{13}e^{-i\dcp} \\
0  &1 &0 \\
-s_{13}e^{i\dcp}  &0 &c_{13} \\
\end{array}
\right)
\left(
\begin{array}{ccc}
c_{12}  &s_{12} &0 \\
-s_{12}  &c_{12} &0 \\
0  &0 &1 \\
\end{array}
\right)
  \nonumber\\
&{\ }& \hspace{-6mm}
=\left(
\begin{array}{ccc}
c_{12}c_{13} & s_{12}c_{13} &  s_{13}e^{-i\dcp}\\
-s_{12}c_{23}-c_{12}s_{23}s_{13}e^{i\dcp} &
c_{12}c_{23}-s_{12}s_{23}s_{13}e^{i\dcp} & s_{23}c_{13}\\
s_{12}s_{23}-c_{12}c_{23}s_{13}e^{i\dcp} &
-c_{12}s_{23}-s_{12}c_{23}s_{13}e^{i\dcp} & c_{23}c_{13}\\
\end{array}
\right)\,,
\label{mns}
\end{eqnarray}
where $c_{jk}\equiv\cos\theta_{jk}$, $s_{jk}\equiv\sin\theta_{jk}$, $\theta_{12}$, $\theta_{23}$, $\theta_{13}$ are three mixing angles, and $\dcp$ is a CP phase. The oscillation probability is expressed based on the elements $U_{\alpha j}~(\alpha= e, \mu, \tau; j= 1,2,3)$ of this mixing matrix and the density of the matter through which the neutrino beam is traveling. Eq.\,(\ref{mns})  shows that $\dcp$ appearance is always accompanied by $\sin\theta_{13}$, which is small; therefore determining $\dcp$ requires precise measurements of the neutrino oscillation probabilities.

In such precision measurements, we must further investigate the effect of the uncertainty in the Earth's density. Refs.\cite{Koike:2000jf,Burguet-Castell:2001ppm,Pinney:2001xw,Huber:2002mx,Ohlsson:2003ip,Huber:2006wb} studied the effect of uncertainty in the density on the oscillation probability measurement, although  mainly in the context of a neutrino factory.  In this paper, we discuss the effect of the uncertainty in the density on  the measurements of oscillation parameters at T2HK and DUNE and their combination.  At these two experiments, where the peak neutrino energy is less than 4 GeV, the matter effect is smaller than that of the mass squared difference term, and therefore the uncertainty in the density is not expected to be as severe as it is at a neutrino factory where the neutrino energy can reach 12 GeV for which causes the matter effect to attain the same magnitude as that of the mass squared difference term.  Since the precision of the measured oscillation parameters has improved recently, it is crucial to study the effect of uncertainty in the density. Ref.\!\cite{Fukasawa:2016yue} previously discussed the sensitivity  to the oscillation parameters of T2HK and DUNE and their combination\footnote{For other phenomenological studies in the context of T2HK and DUNE in the standard three flavor scenario, we refer to Refs.~\cite{Ghosh:2017ged,Chatterjee:2017xkb,Blennow:2014sja,Chakraborty:2018dew,Chakraborty:2017ccm,Agarwalla:2017wct,Blennow:2020ncm,Coloma:2012wq,Coloma:2012ji,Ballett:2016daj}}. Our motivation for updating the findings of Ref.\!\cite{Fukasawa:2016yue} is two fold as follows: (i) After Ref.\!\cite{Fukasawa:2016yue} was published, the setup of the T2HK and DUNE experiments were updated. Regarding T2HK, only one talk at the far detector site is assumed in the first phase; (ii) The uncertainty in the Earth's density was not considered in the analysis Ref.\!\cite{Fukasawa:2016yue}.\footnote{The  effect of the uncertainty in the Earth's density on the oscillation probabilities at T2HK and DUNE was studied in Refs.\!\cite{Kelly:2018kmb,King:2020ydu,Schwetz:2021thj}. But they do not study the effect of the Earth's density on the measurement of the unknown neutrino oscillation parameters.} As we will observe in subsequent sections, the uncertainty in the density gives a non-negligible contribution to the precision of the
the CP phase.

Unlike in Ref.\!\cite{Fukasawa:2016yue}, we  did not include the atmospheric neutrino measurement at Hyper-kamiokande, because a treatment of the uncertainty in the density is quite complicated for the analysis of atmospheric neutrinos. For simplicity, we assume that the true mass ordering is normal
ordering.

This paper is organized as follows. In sect. \ref{Parameter degeneracy}, we briefly introduce the parameter degeneracy. In sect. \ref{Analysis}, we describe basic parameters of the T2HK and DUNE experiments, and comprehensively discuss our simulation details, and  in sect. \ref{Results}, we  present the results of our analysis. Furthermore, we summarize our conclusions in sect. \ref{conclusion}. Moreover, in the appendix \ref{appendixa}, we  also provide a discussion on the octant degeneracy which appears at DUNE.

\section{Parameter degeneracy}
\label{Parameter degeneracy}

In research of neutrino oscillation with the standard three flavor scenario, determination of the CP phase $\dcp$ is regarded as the important goal.  It has been known that so-called parameter degeneracy makes it difficult for us to determine $\dcp$ uniquely, even if we are given the values of the appearance oscillation probabilities $P(\nu_\mu\to\nu_e)$ and $P(\bar{\nu}_\mu\to\bar{\nu}_e)$ at a fixed neutrino energy and at a fixed baseline length.

Before $\theta_{13}$ was precisely measured by the reactor neutrino experiments\cite{Workman:2022ynf}, three kinds of parameter degeneracy existed. The first was the intrinsic degeneracy\!\cite{Burguet-Castell:2001ppm}, which occurs since the appearance oscillation probabilities are approximately quadratic in $\sin2\theta_{13}$ for small $\theta_{13}$; thus it yields the two solutions $(\theta_{13}, \dcp)$ and $(\theta_{13}^\prime, \dcp^\prime)$. The second is the sign degeneracy\!\cite{Minakata:2001qm}. The value of $\dcp$ depends on whether the true mass ordering is normal or inverted; therefore, determining mass hierarchy is  crucial. The third one is the octant degeneracy\!\cite{Fogli:1996pv}. The primary contribution to the probabilities $1-P(\nu_\mu\to\nu_\mu)$ and $1-P(\bar{\nu}_\mu\to\bar{\nu}_\mu)$ is proportional to $\sin^22\theta_{23}$, and we can only determine $\sin^22\theta_{23}$ from the disappearance channel; subsequently,  if $\theta_{23}$ is not maximal, then we have two solutions $(\theta_{23}, \dcp)$ and $(90^\circ-\theta_{23}, \dcp^\prime)$.

These three types of degeneracies together created an eight fold degeneracy \cite{Barger:2001yr} when the precise value of $\theta_{13}$ was unknown. After the precise value of $\theta_{13}$ was determined by the reactor neutrino experiments\,\cite{Workman:2022ynf}, some of the eight fold
degeneracy is resolved. However, in a practical experimental situation, the formal discussions in Ref.\!\cite{Burguet-Castell:2001ppm,Minakata:2001qm,Fogli:1996pv} $-$ in which the probabilities $P(\nu_\mu\to\nu_e)$ and $P(\bar{\nu}_\mu\to\bar{\nu}_e)$ are assumed to be determined precisely $-$ may not necessarily apply to the T2HK or DUNE experiments because of experimental errors.  Since the cross sections for neutrinos and antineutrinos vary, the number of events (and therefore the statistical errors) for neutrinos and antineutrinos can differ. Previously, some authors discussed parameter degeneracy by investigating only the neutrino mode, and degeneracy which occurs in the neutrino oscillation probability is known as the ``generalized hierarchy-octant-$\dcp$ degeneracy''
\cite{Ghosh:2015ena}.  This generalized degeneracy comprises hierarchy-$\dcp$ degeneracy \cite{Prakash:2012az} and octant-$\dcp$ degeneracy \cite{Agarwalla:2013ju}. When the number of events for antineutrinos is not sufficiently large, ``generalized hierarchy-octant-$\dcp$ degeneracy''
becomes relevant. Note that when T2HK and DUNE will be running, the medium baseline reactor experiment JUNO \cite{JUNO:2015zny} may provide a clear answer of the neutrino mass ordering. In that case the degeneracy associated with the mass ordering will be resolved.

\section{Analysis}
\label{Analysis}
\subsection{T2HK and DUNE}

T2HK is the long baseline experiment which is planned in Japan, and its baseline length and peak energy is $L$=295 km, $E\sim 0.6$ GeV, respectively. In the simulation of T2HK, we follow the configuration as given in Ref.\!\cite{Hyper-Kamiokande:2016srs}. We consider one water-Cerenkov detector tank having fiducial volume of 187~kt located at Kamioka which is 295 km from the neutrino source at J-PARC having a beam power of 1.3 MW with a total exposure of $27 \times 10^{21}$ protons on target, corresponding to 10~years of running. We have divided the total run-time into 5 years in neutrino mode and 5 years in anti-neutrino mode. This assumption differs from the original setup in Ref.\!\cite{Hyper-Kamiokande:2016srs} where  a 3:1 ratio of antineutrino mode to neutrino mode operation is assumed.\footnote{The reason why we work with 1:1 ratio rather than 3:1 is because it is more advantageous for T2K to run in the dominant neutrino mode than to run with the ratio of 3:1, where the antineutrino mode is required only in resolving the octant degeneracy, as was described by an author in Ref.\!\cite{Ghosh:2015tan}.} The reference value for the Earth's density is 2.70~g/cm$^3$, as given in Ref.\!\cite{Hyper-Kamiokande:2016srs}.
  
DUNE is another long baseline experiment which is planned in USA, and its baseline length and peak energy is $L$=1300 km, $E\sim3$ GeV, respectively. In the case of DUNE, we have used the official GLoBES files of the DUNE technical design report~\cite{DUNE:2021cuw} which reproduces the results presented in Ref.\!\cite{DUNE:2020ypp}. A 40~kt liquid argon time-projection chamber detector is placed 1300~km from the source having a power of 1.2~MW delivering $1.1 \times 10^{21}$ protons on target per year with a running time of 7~years. We have divided the run-time into 3.5 years in neutrino mode and 3.5 years in antineutrino mode. The 1:1 ratio of antineutrino mode to neutrino mode is the same setup as that assumed in Ref.\!\cite{DUNE:2021cuw}. The neutrino source will be located at Fermilab, USA and the detector will be located at South Dakota, USA.
The reference value for the Earth's density is 2.848~g/cm$^3$, as given in Ref.\!\cite{DUNE:2021cuw}.

This work is an update of the previous work\cite{Fukasawa:2016yue} and adopted here the latest configuration of each experiment, including
the detector volume, beam power, event selection, among others. In this study, we employ the configuration of T2HK from Ref.\!\cite{Hyper-Kamiokande:2016srs} and that of DUNE from Ref.\!\cite{DUNE:2020ypp}, whereas in the previous work\cite{Fukasawa:2016yue}
we took the T2HK configuration from Ref.\!\cite{Hyper-KamiokandeWorkingGroup:2014czz} and that of DUNE from Ref.\!\cite{DUNE:2015lol}. Some  preliminary results from this work were presented by an author at Neutrino 2022\cite{Yasuda:2022nu}. After then, we made an additional effort to match our event spectrum with the one in Ref.\!\cite{Hyper-Kamiokande:2016srs} by tweaking the systematics a little to reproduce the results in Ref.\!\cite{Hyper-Kamiokande:2016srs} more accurately.

\subsection{Simulation Details}

The experiments T2HK and DUNE are simulated using the software GLoBES \cite{Huber:2004ka,Huber:2007ji}. We perform our analysis with the Poisson log-likelihood function $\chi^2$, which depends on the two sets of the oscillation parameters and the Earth's density $\rho$ 
$\vec{p}_{\mbox{\rm\scriptsize ex}}=(\Delta m^2_{21}, \Delta m^2_{31}, \theta_{12}, \theta_{13}, \theta_{23}, \dcp, \rho)_{\mbox{\rm\scriptsize true}}$
for the "true" parameters and $\vec{p}_{\mbox{\rm\scriptsize th}}=(\Delta m^2_{21}, \Delta m^2_{31}, \theta_{12}, \theta_{13}, \theta_{23}, \dcp, \rho)_{\mbox{\rm\scriptsize test}}$ for the "test" ones:
\begin{eqnarray}
&{\ }& \hspace{-15mm} \chi^2 (\vec{p}_{\mbox{\rm\scriptsize
ex}},\vec{p}_{\mbox{\rm\scriptsize th}}) =
\min_{\{\xi_j^{(\alpha)}\}}
\bigg[ \chi^2(\vec{p}_{\mbox{\rm\scriptsize ex}},
\vec{p}_{\mbox{\rm\scriptsize th}}; \{\xi_j^{(\alpha)}\})
%    \nonumber\\
%    &{\ }& \hspace{-0mm}
+ {\displaystyle \sum_{\alpha=e,\mu}\sum_{j=1}^4}
\left(\frac{\xi_j^{(\alpha)}}{\pi_j^{(\alpha)}}\right)^2 \bigg],
\label{chi2v0}
\end{eqnarray}
where $\{ \xi_j^{(\alpha)} \}$ are the pull variables, 
$\pi_j^{(\alpha)}$ are the ($1\sigma$) systematic errors for the pull variable $\xi_j^{(\alpha)}$, 
\begin{eqnarray}
&{\ }& \hspace{-20mm}
\chi^{2 \ }
(\vec{p}_{\mbox{\rm\scriptsize ex}},
\vec{p}_{\mbox{\rm\scriptsize th}};
\{\xi_j^{(\alpha)}\}) = 2 {\displaystyle
\sum_{\alpha=e,\mu}\sum_i} \Bigg[
M_{i}^{(\alpha)}(\vec{p}_{\mbox{\rm\scriptsize th}}) -
N_{i}^{(\alpha)}(\vec{p}_{\mbox{\rm\scriptsize ex}})
\nonumber\\
&{\ }& \hspace{35mm}
+N_{i}^{(\alpha)}(\vec{p}_{\mbox{\rm\scriptsize ex}}) \ln \left(
{\displaystyle \frac{N_{i}^{(\alpha)}(\vec{p}_{\mbox{\rm\scriptsize
ex}})} {M_{i}^{(\alpha)}(\vec{p}_{\mbox{\rm\scriptsize
th}})} } \right) \Bigg]
\label{chi2v1}
\end{eqnarray}
is a $\chi^2$ function defined from the Poisson statistics,
the "experimental" $\alpha$-like data
$N_{i}^{(\alpha)}(\vec{p}_{\mbox{\rm\scriptsize ex}})~(\alpha=e,\mu)$
for the $i^{\rm th}$ energy bin
are defined as
the sum of the numbers of events for signals ($S_{i}^{(\alpha)}$) and
for backgrounds ($B_{i}^{(\alpha)}$)
\begin{eqnarray*}
&{\ }& \hspace{-50mm}
N_{i}^{(\alpha)} (\vec{p}_{\mbox{\rm\scriptsize ex}})
= S_{i}^{(\alpha)} (\vec{p}_{\mbox{\rm\scriptsize ex}})
+ B_{i}^{(\alpha)} (\vec{p}_{\mbox{\rm\scriptsize ex}})\,,
\nonumber
\end{eqnarray*}
the "theoretical" $\alpha$-like events
$M_{i}^{(\alpha)}(\vec{p}_{\mbox{\rm\scriptsize th}})$ are defined as
\begin{eqnarray}
&{\ }& \hspace{-20mm}
M_{i}^{(\alpha)} (\vec{p}_{\mbox{\rm\scriptsize th}})
= \left( 1 + \xi_1^{(\alpha)} + \xi_3^{(\alpha)}
\frac{E_i - E_{\rm av}}{E_{\rm max} - E_{\rm min
}} 
\right) S_{i}^{(\alpha)} (\vec{p}_{\mbox{\rm\scriptsize th}})
\nonumber\\
&{\ }& \hspace{-0mm}
+ \left( 1 + \xi_2^{(\alpha)} + \xi_4^{(\alpha)}
\frac{E_i - E_{\rm av}}{E_{\rm max} - E_{\rm min
}}\right) B_{i}^{(\alpha)} (\vec{p}_{\mbox{\rm\scriptsize th}})\,
\label{chi2v2}
\end{eqnarray}
using the "test" oscillation parameters $\vec{p}_{\mbox{\rm\scriptsize th}}$ together with the systematic uncertainty $\xi_j^{(\alpha)}$.  In Eq.\,(\ref{chi2v1}) the subscript $i$ runs over all the energy bins in both the appearance ($e$-like) and disappearance ($\mu$-like) channels.
Table~\ref{table_sys} presents the systematic errors used in our calculations for the two experiments.  $\pi_j^{(\alpha)}~(j=1,2)$ ($\pi_j^{(\alpha)}~(j=3,4)$) corresponds to the relevant normalization (tilt) systematic errors for a given experimental observable.  For our simulation we have fixed the tilt error $\pi_j^{(\alpha)}~(j=3,4)$ to a constant value which is 10\% for T2HK\footnote{The values for the tilt variables $\xi_j^{(\alpha)}~(j=3,4)$ are chosen such that our simulation results for T2HK match with the sensitivities reported in the collaboration study.} and 2.5\% for DUNE corresponding to all the channels.  In Eq.\,(\ref{chi2v2}) $E_i$ is the mean energy of the $i^{\rm th}$ energy bin, $E_{\mbox{\rm\scriptsize min}}$ and $E_{\mbox{\rm\scriptsize max}}$ are the limits of the full energy range, and $E_{\mbox{\rm\scriptsize av}}$ is their average.  The normalization errors affect the scaling of events and the tilt errors influence the energy dependence of the events.  All the pull variables $\{ \xi_j^{(\alpha)} \}$ take values in the range $(-3\pi_j^{(\alpha)},3\pi_j^{(\alpha)})$, so that the errors can vary from $-3\sigma$ to $+3\sigma$.
$\chi^2(\vec{p}_{\mbox{\rm\scriptsize ex}},\vec{p}_{\mbox{\rm\scriptsize th}})$ is then calculated  by minimizing over all combinations of $\xi_j^{(\alpha)}$.

\begin{table} [h]
\centering
\begin{tabular}{|c|c|c|} \hline
Systematics            & T2HK               & DUNE  \\ \hline
Sg-norm $\nu_{e}$  & 4.71\% (4.47\%)   & 2\% (2\%)      \\ 
Sg-norm $\nu_{\mu}$ & 4.13\% (4.15\%)      & 5\%  (5\%)\\ 
Bg-norm $\nu_e$ & 4.71\% (4.47\%)             & 5\% to 20\% (5\% to 20\%)\\ 
Bg-norm $\nu_\mu$ & 4.13\% (4.15\%)          & 5\% to 20\% (5\% to 20\%) \\ 
\hline
\end{tabular}
\caption{The values of systematic errors that we considered in our analysis. ``norm" denotes the normalization error, ``Sg" represents the signal and ``Bg" signifies the background. The numbers without (with) parenthesis are for neutrinos (antineutrinos). For T2HK, the systematics are the same for signal and background whereas for DUNE the systematics errors are the same for neutrinos and antineutrinos.}
\label{table_sys}
\end{table}

\begin{table}
\centering
\begin{tabular}{|c|c|c|} \hline
Parameters            & True values               & Test value Range  \\ \hline
$\sin^2 \theta_{12}$  & $33.45^\circ$ & Fixed      \\ 
$\sin^2 \theta_{13}$ & $8.62^\circ$                     & Fixed \\ 
$\sin^2 \theta_{23} $ & $42^\circ$/$48^\circ$                  & $39^\circ$ - $51^\circ$\\ 
$\dcp $       & $ -90^\circ$/$0^\circ$                  & $-180^\circ$ - $180^\circ $\\ 
$\Delta m^2_{12}$    & $7.42 \times 10^{-5}~{\rm eV}^2 $ & Fixed \\ 
$\Delta m^2_{31}$    &~ $ 2.510 \times 10 ^{-3}~{\rm  eV}^2$ (NO)~~& $2.43 \times 10 ^{-3}~{\rm  eV}^2$ - $2.60 \times 10 ^{-3}~{\rm  eV}^2$ \\
density (T2HK) & $\rho_0$=2.70~g/cm$^3$ & $|\rho/\rho_0-1|\le 0.05$
,~$|\rho/\rho_0-1|\le 0.1$\\
density (DUNE) & $\rho_0$=2.848~g/cm$^3$ & $|\rho/\rho_0-1|\le 0.05$
,~$|\rho/\rho_0-1|\le 0.1$\\
\hline
\end{tabular}
\caption{The true values of the oscillation parameters and the Earth's density and the range of the test variables that we considered in our analysis. The label ``NO'' represents the normal ordering of the neutrino masses.}
\label{table_param}
\end{table}    

To obtain sensitivity to an oscillation parameter including $\dcp$, we marginalize $\chi^2 (\vec{p}_{\mbox{\rm\scriptsize ex}},\vec{p}_{\mbox{\rm\scriptsize th}})$ in Eq.\,(\ref{chi2v0}), i.e., we minimize it with respect to the test parameters $\vec{p}_{\mbox{\rm\scriptsize th}}$ including the Earth's density. The true values of the oscillation parameters and the range of the test ones are taken from Nufit v5.1 \cite{Esteban:2020cvm} and they are presented together with the values of the density in Table \ref{table_param}. As for the the uncertainty in the Earth's density, the reasonable
reference value for the uncertainty in the density is 5\% based on Ref.\!\cite{Geller:2001ix}. However, we also analyzed a more conservative case
for which the the uncertainty is 10\%. In our analysis, we do not introduce priors for the test oscillation parameters or for the test variable for the  Earth's density, over which we marginalize.

\section{Results}
\label{Results}

\subsection{Sensitivity to mass ordering}
%Fig.1
\begin{figure}[H]
\begin{center}
\includegraphics[scale=0.42]{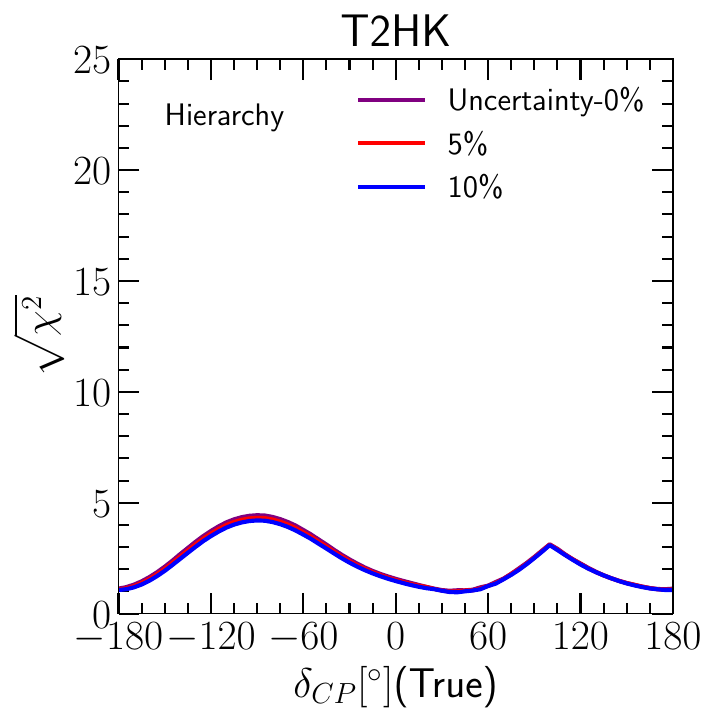}
\includegraphics[scale=0.42]{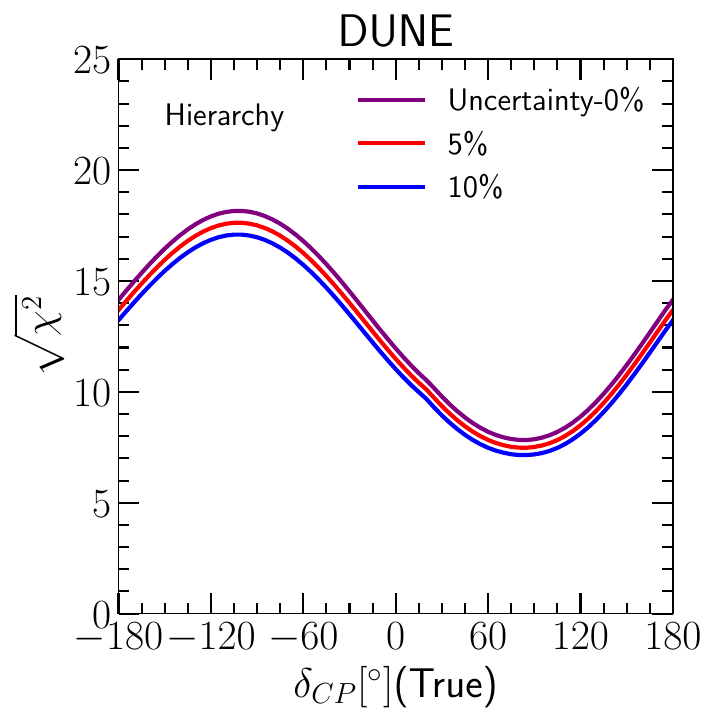}     
\includegraphics[scale=0.42]{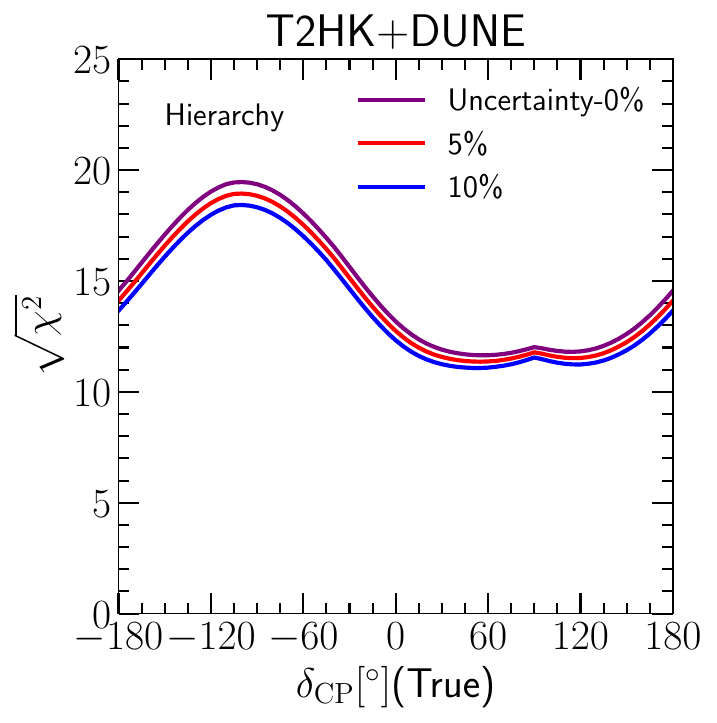}
\caption{Mass ordering sensitivity as a function of $\dcp$
(true). $\theta_{23} = 42^\circ$.}
\label{hier}
\end{center}
\end{figure}

Sensitivity to mass ordering can be expressed by $\chi^2$ as a function of true $\dcp$ for each experiment, where $\chi^2$ is defined by
\begin{eqnarray}
%\hspace*{-10mm}
&{\ }& \hspace{-20mm}
\chi^2_{\mbox{\rm\scriptsize T2K}} \equiv
\min_{\mbox{\rm\scriptsize param.}}\min_\rho~
\chi^2_{\mbox{\rm\scriptsize T2K}} (\vec{p}_{\mbox{\rm\scriptsize ex}}(NO),
\vec{p}_{\mbox{\rm\scriptsize th}}(IO))
\label{hierarchy-chi2-t2k}\\
&{\ }& \hspace{-20mm}
\chi^2_{\mbox{\rm\scriptsize DUNE}} \equiv
\min_{\mbox{\rm\scriptsize param.}}\min_\rho~
\chi^2_{\mbox{\rm\scriptsize DUNE}} (\vec{p}_{\mbox{\rm\scriptsize ex}}(NO),
\vec{p}_{\mbox{\rm\scriptsize th}}(IO))
\label{hierarchy-chi2-dune}\\
&{\ }& \hspace{-20mm}
\chi^2_{\mbox{\rm\scriptsize combined}} \equiv
\min_{\mbox{\rm\scriptsize param.}}\left[
\min_\rho~
\chi^2_{\mbox{\rm\scriptsize T2K}} (\vec{p}_{\mbox{\rm\scriptsize ex}}(NO),
\vec{p}_{\mbox{\rm\scriptsize th}}(IO))\right.
\nonumber\\
&{\ }& \hspace{20mm}
\left.+\min_\rho~
\chi^2_{\mbox{\rm\scriptsize DUNE}} (\vec{p}_{\mbox{\rm\scriptsize ex}}(NO),
\vec{p}_{\mbox{\rm\scriptsize th}}(IO))\right]
\label{hierarchy-chi2-comb}
\end{eqnarray}
where $\chi^2 (\vec{p}_{\mbox{\rm\scriptsize ex}}(NO), \vec{p}_{\mbox{\rm\scriptsize th}}(IO))$ is defined in Eq.\,(\ref{chi2v0}), ``param.'' represents
the test oscillation parameters ($\theta_{23}$, $\dcp$, $|\Delta m^2_{31}|$, sign($\Delta m^2_{31}$)) over which we are marginalizing, NO (IO) represents normal (inverted) ordering, respectively, and $\rho$ is the test density variable of the Earth. Since measurements of the density for the T2HK and DUNE sites are presumed to be independent, we marginalize over each test density separately.

\begin{figure}[H]
\begin{center}
\includegraphics[scale=0.42]{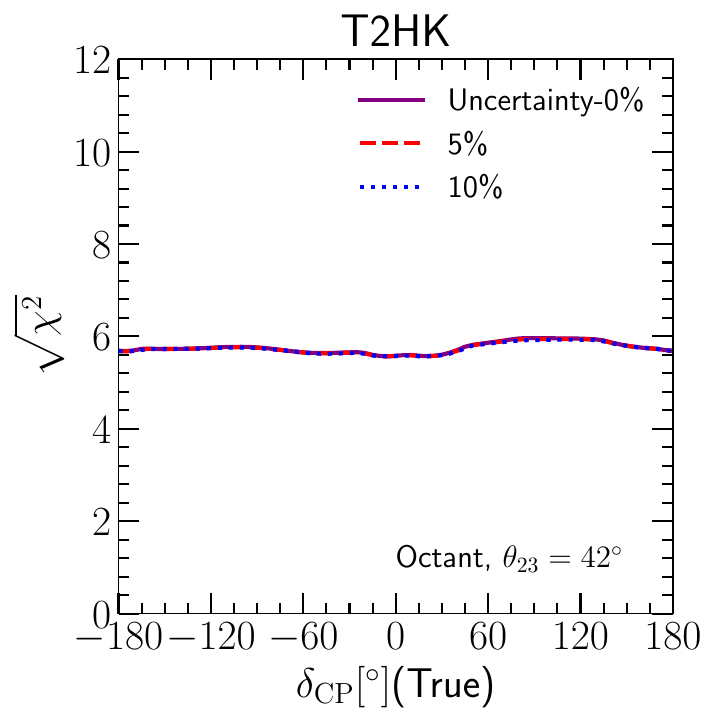}
\includegraphics[scale=0.42]{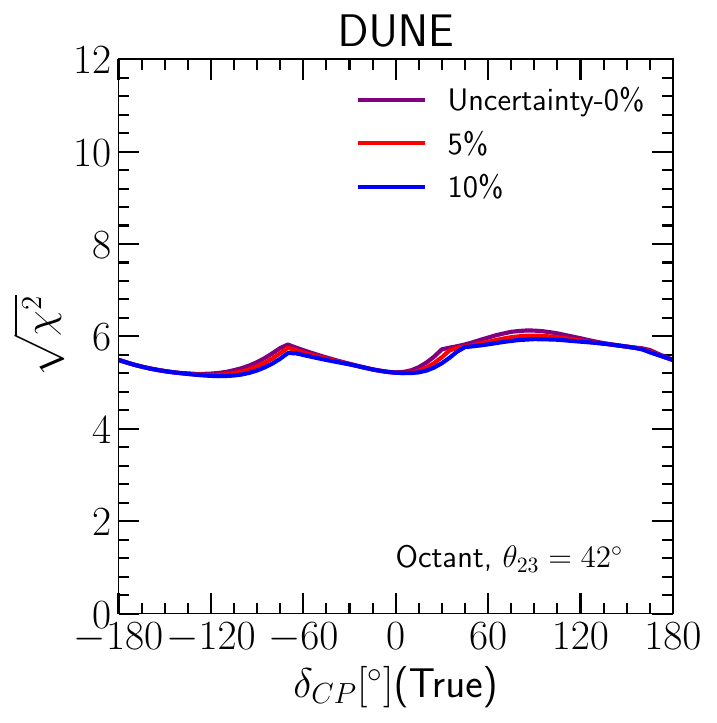}     
\includegraphics[scale=0.42]{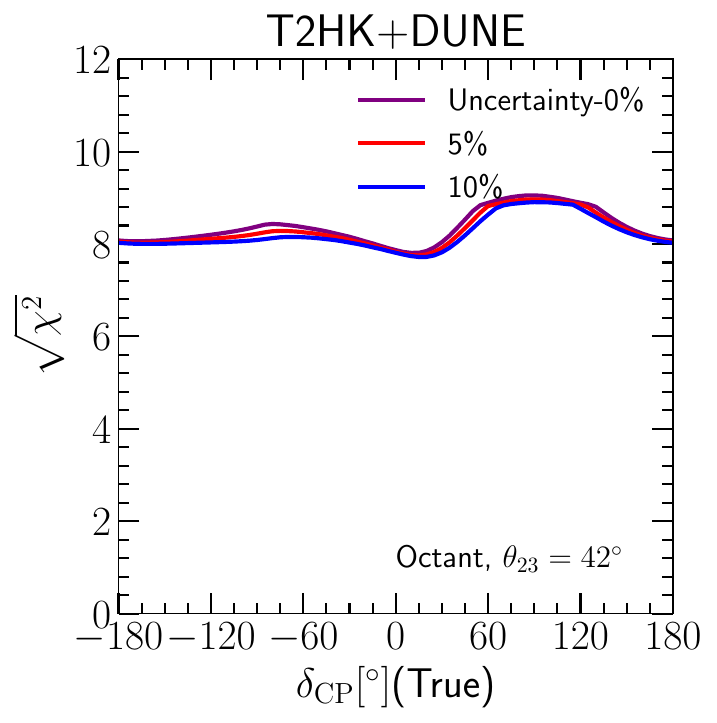} \\
\includegraphics[scale=0.42]{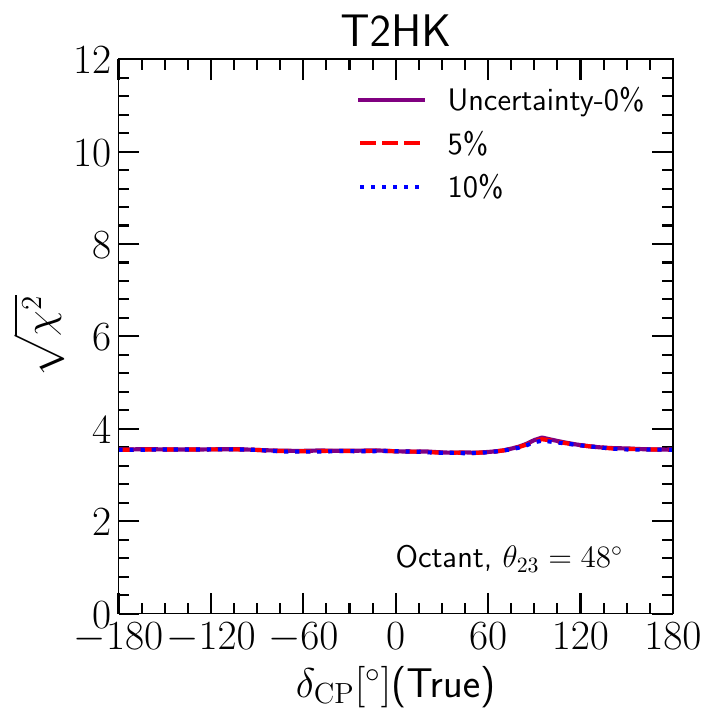}
\includegraphics[scale=0.42]{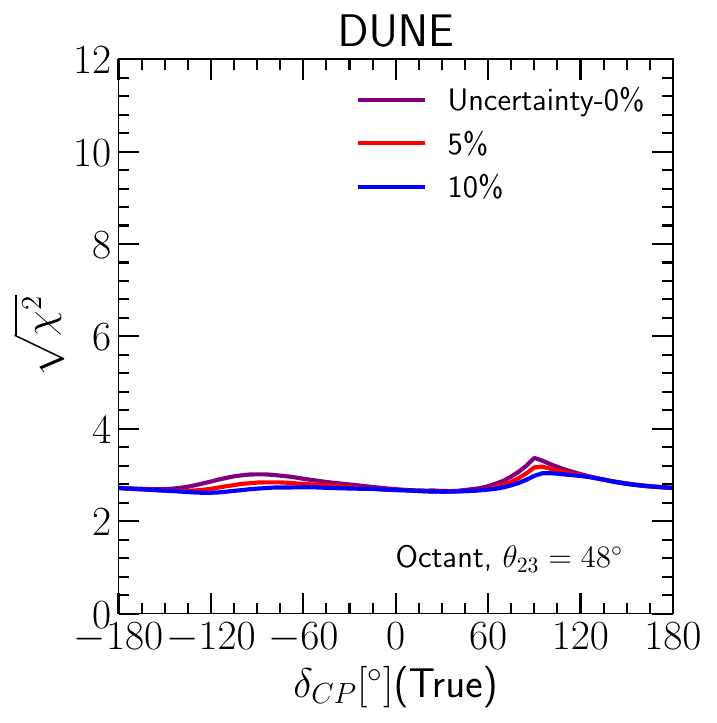}     
\includegraphics[scale=0.42]{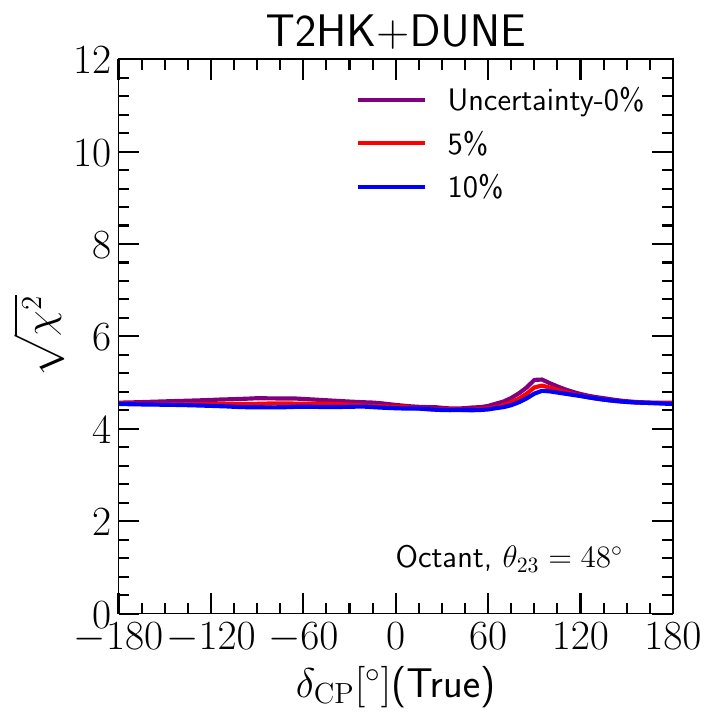} \\
\includegraphics[scale=0.43]{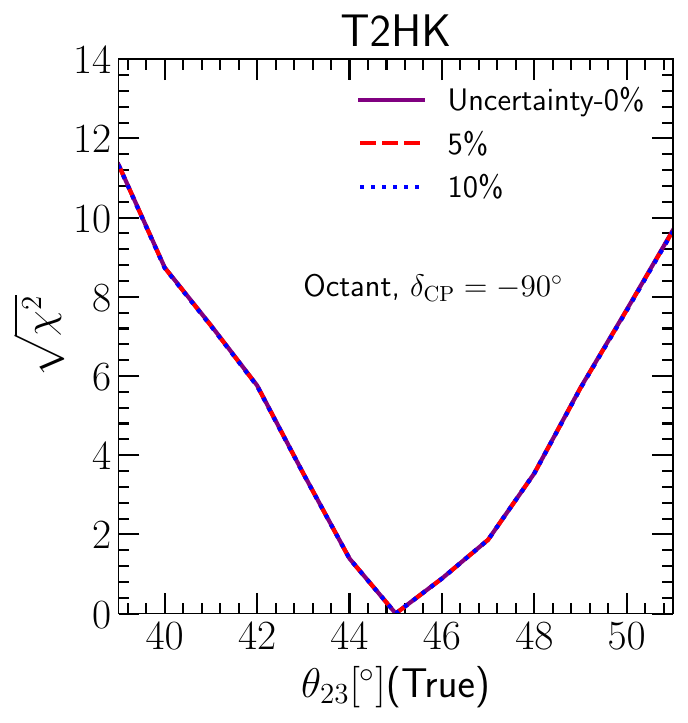}
\includegraphics[scale=0.43]{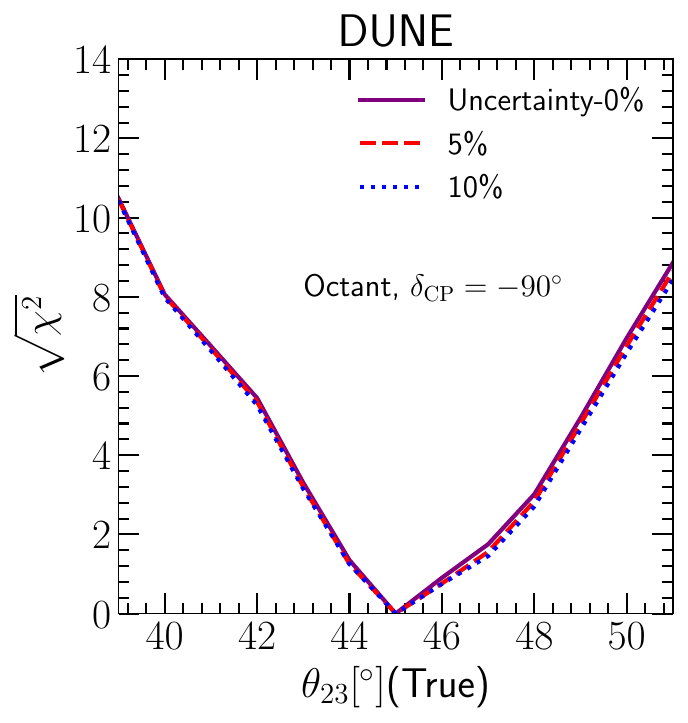}     
\includegraphics[scale=0.43]{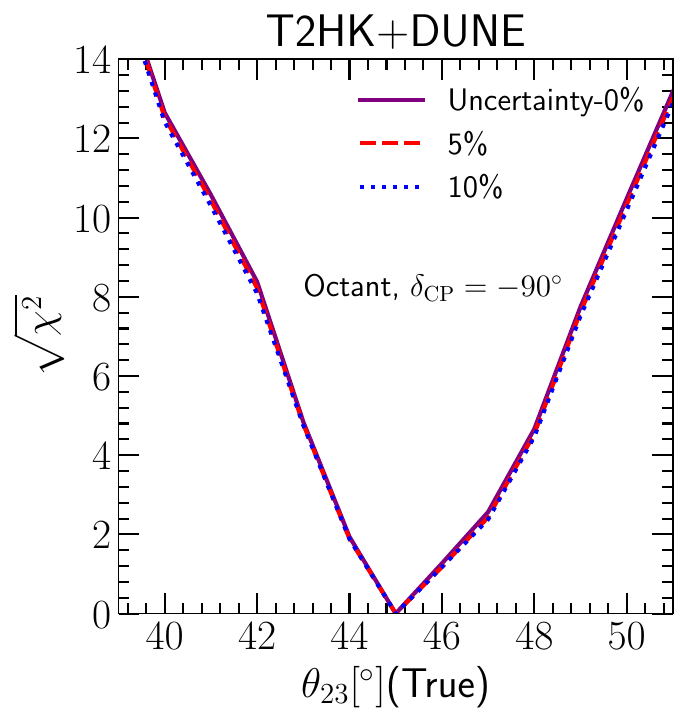}
\caption{Octant sensitivity as a function of $\dcp$ (true)
(top and middle panel) and $\theta_{23}$ true (bottom
panel). Hierarchy unknown.}
\label{octa}
\end{center}
\end{figure}

The square root of $\chi^2$, which stands for significance to reject wrong mass ordering, is plotted as a function of the true value of $\dcp$ in Fig.\ref{hier} for three different reference values (0\%, 5\%, and 10\%) for the uncertainty in the test density variable $\rho$. The left panel is for T2HK, the middle panel is for DUNE and the right panel is for the combination of T2HK and DUNE, denoted by T2HK+DUNE. The true value of $\theta_{23}$ is $42^\circ$. From Fig.\ref{hier}, we conclude that, irrespective of the value of true $\dcp$, significance to reject wrong mass ordering is 1, 8 and 11 for
T2HK, DUNE and their combination, respectively. As we can see from Fig.\ref{hier}, the significance is small for T2HK whereas it is quite large for DUNE and combination of the two. Therefore, the uncertainty in the density affects the result of DUNE more than it does that of T2HK, as is naively expected from the length of their baselines.

We note in passing that significance for mass ordering in the combination of T2HK and DUNE for $\dcp \sim 90^\circ$ is enhanced because of synergy.
From mass order degeneracy, in principle we expect two possible values for $\dcp$ at each experiment. For DUNE, this ambiguity is removed because of its long baseline; thus the DUNE data provide us a unique choice for $\dcp$, as far as the mass order degeneracy is concerned. If we apply this information to T2HK, then the ambiguity at T2HK is removed and the contribution of T2HK to significance is enhanced.

%\vglue 8cm
%\newpage
\subsection{Sensitivity to octant degeneracy}

The sensitivity to the octant is defined by
\begin{eqnarray}
%\hspace*{-10mm}
&{\ }& \hspace{-20mm}
\chi^2_{\mbox{\rm\scriptsize T2K}} \equiv
\min_{\mbox{\rm\scriptsize param.}}\min_\rho~
\chi^2_{\mbox{\rm\scriptsize T2K}} (\vec{p}_{\mbox{\rm\scriptsize ex}}(RO),
\vec{p}_{\mbox{\rm\scriptsize th}}(WO))
\label{octant-chi2-t2k}\\
&{\ }& \hspace{-20mm}
\chi^2_{\mbox{\rm\scriptsize DUNE}} \equiv
\min_{\mbox{\rm\scriptsize param.}}\min_\rho~
\chi^2_{\mbox{\rm\scriptsize DUNE}} (\vec{p}_{\mbox{\rm\scriptsize ex}}(RO),
\vec{p}_{\mbox{\rm\scriptsize th}}(WO))
\label{octant-chi2-dune}\\
&{\ }& \hspace{-20mm}
\chi^2_{\mbox{\rm\scriptsize combined}} \equiv
\min_{\mbox{\rm\scriptsize param.}}\left[
\min_\rho~
\chi^2_{\mbox{\rm\scriptsize T2K}} (\vec{p}_{\mbox{\rm\scriptsize ex}}(RO),
\vec{p}_{\mbox{\rm\scriptsize th}}(WO))\right.
\nonumber\\
&{\ }& \hspace{10mm}
\left.+\min_\rho~
\chi^2_{\mbox{\rm\scriptsize DUNE}} (\vec{p}_{\mbox{\rm\scriptsize ex}}(RO),
\vec{p}_{\mbox{\rm\scriptsize th}}(WO))\right]
\label{octant-chi2-comb}
\end{eqnarray}
%In Eq.\,(\ref{octant-chi2})
where ``RO'' represents the right octant, ``WO'' represents the wrong octant and ``param.'' represents the test variables ($\dcp$, $|\Delta m^2_{31}|$, sign($\Delta m^2_{31}$)) which we are marginalizing over. 

In the top (middle) panel of Fig.\ref{octa}, significance is depicted as a function of true value of $\dcp$ for the true value $\theta_{23}=42^\circ$ ($\theta_{23}=48^\circ$) for three different reference values (0\%, 5\%, and 10\%) for the uncertainty in the test density variable $\rho$. In each row, the left panel is for T2HK, the middle panel is for DUNE and the right panel is for the combination of T2HK and DUNE. From these panels we see that significance to establish the correct octant for LO (HO) is 5.5 (3), 5 (2.5) and 7 (4.5) for T2HK, DUNE and their combination, respectively. Note that the octant sensitivity in the LO is in general higher as compared to HO because of the higher number in denominator of the $\chi^2$ distribution formula.
In the bottom panel of Fig.\ref{octa}, significance is shown as a function of true value of $\theta_{23}$, where the true value of $\dcp$ is assumed to be
$-90^\circ$.  From this we observe that octant degeneracy can be resolved at $5 \sigma$ C.L unless $42.3^\circ < \theta_{23} < 48.7^\circ$ for T2HK, $42.2^\circ < \theta_{23} < 49.1^\circ$ for DUNE, or $42.9^\circ < \theta_{23} < 48.0^\circ$ for their combination. In all the plots in Fig.\ref{octa}, the uncertainty in the density is not important.

%\newpage
\subsection{Sensitivity to $\dcp$}

Sensitivity to CP violation is evaluated by
\begin{eqnarray}
%\hspace*{-10mm}
&{\ }& \hspace{-40mm}
\chi^2_{\mbox{\rm\scriptsize T2K}} \equiv
\min_{\mbox{\rm\scriptsize param.}}\,\min_{\dcp^\prime=0,180^\circ}\,\min_\rho~
\chi^2_{\mbox{\rm\scriptsize T2K}} (\vec{p}_{\mbox{\rm\scriptsize ex}},
\vec{p}_{\mbox{\rm\scriptsize th}})
\label{cp-delta-chi2-t2k}\\
&{\ }& \hspace{-40mm}
\chi^2_{\mbox{\rm\scriptsize DUNE}} \equiv
\min_{\mbox{\rm\scriptsize param.}}\,\min_{\dcp^\prime=0,180^\circ}\,\min_\rho~
\chi^2_{\mbox{\rm\scriptsize DUNE}} (\vec{p}_{\mbox{\rm\scriptsize ex}},
\vec{p}_{\mbox{\rm\scriptsize th}})
\label{cp-delta-chi2-dune}\\
&{\ }& \hspace{-40mm}
\chi^2_{\mbox{\rm\scriptsize combined}} \equiv
\min_{\mbox{\rm\scriptsize param.}}\,\min_{\dcp^\prime=0,180^\circ}\left[
\min_\rho~
\chi^2_{\mbox{\rm\scriptsize T2K}} (\vec{p}_{\mbox{\rm\scriptsize ex}},
\vec{p}_{\mbox{\rm\scriptsize th}})\right.
\nonumber\\
&{\ }& \hspace{10mm}
\left.+\min_\rho~
\chi^2_{\mbox{\rm\scriptsize DUNE}} (\vec{p}_{\mbox{\rm\scriptsize ex}},
\vec{p}_{\mbox{\rm\scriptsize th}})\right]
\label{cp-delta-chi2-comb}
\end{eqnarray}
where ``param.'' represents the test oscillation parameters over which we are marginalizing ($\theta_{23}$, $|\Delta m^2_{31}|$, sign($\Delta m^2_{31}$)), and we are also marginalizing over the test CP phase $\dcp^\prime = 0^\circ$ or $180^\circ$, i.e., choosing the smaller value from the two cases $\dcp^\prime = 0^\circ$ or $180^\circ$, as well as over the test density variable $\rho$.

\begin{figure}[H]
\begin{center}
\includegraphics[scale=0.6]{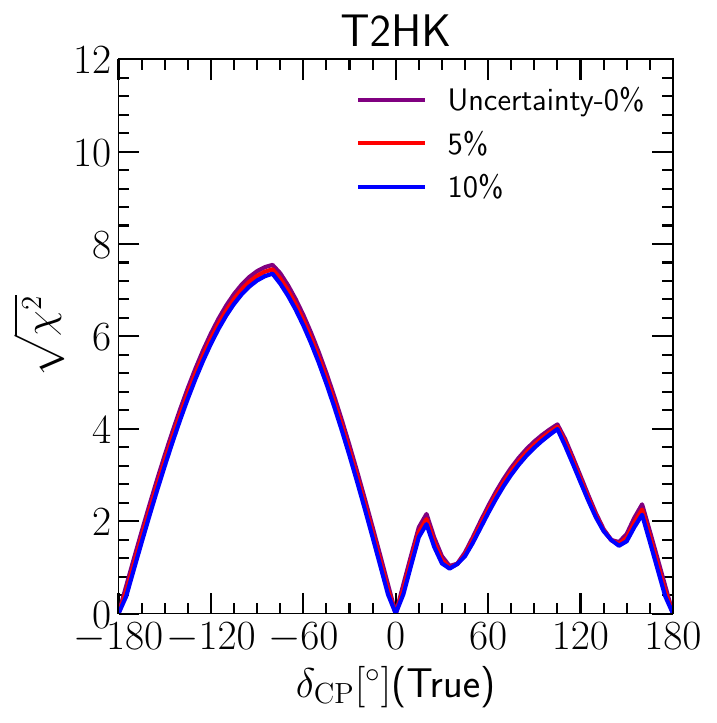}
%\hspace{0.8 in}   
\includegraphics[scale=0.6]{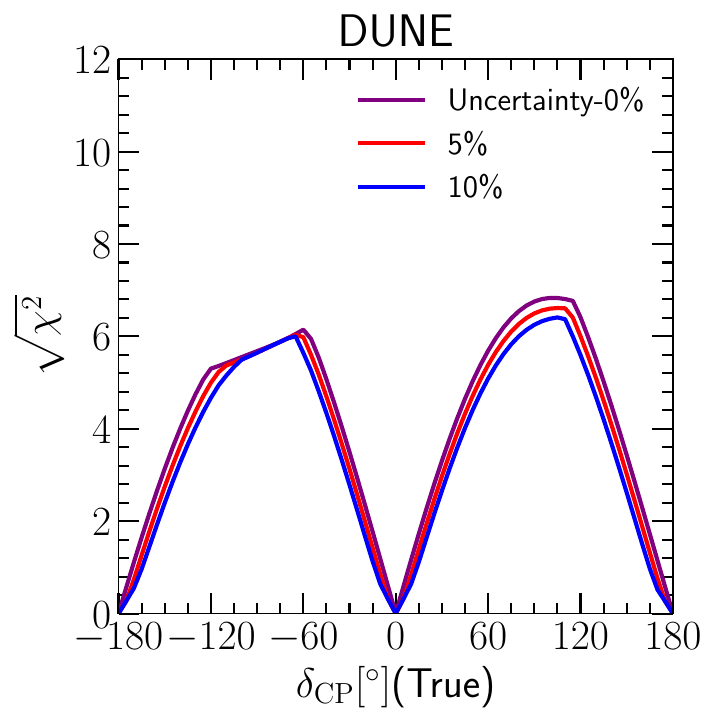} \\     
\includegraphics[scale=0.6]{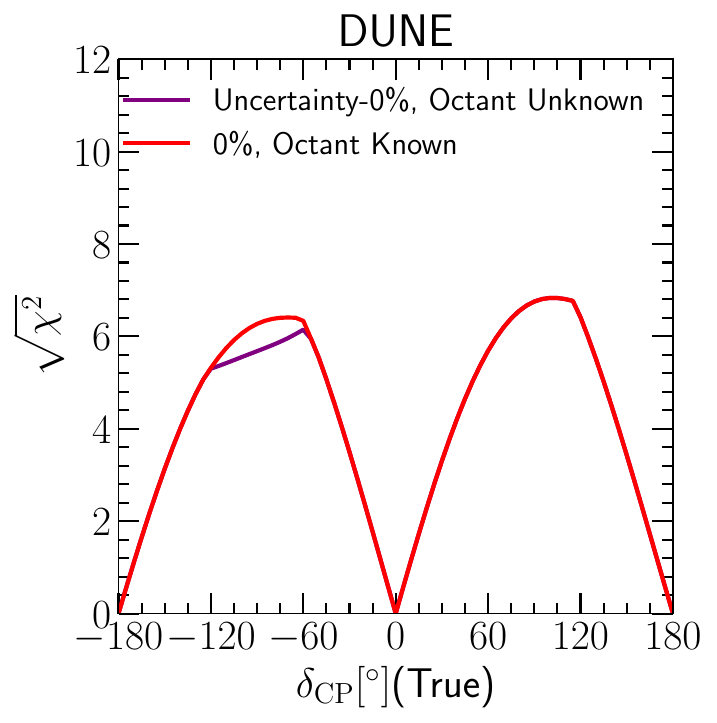}
%\hspace{0.8 in}
\includegraphics[scale=0.6]{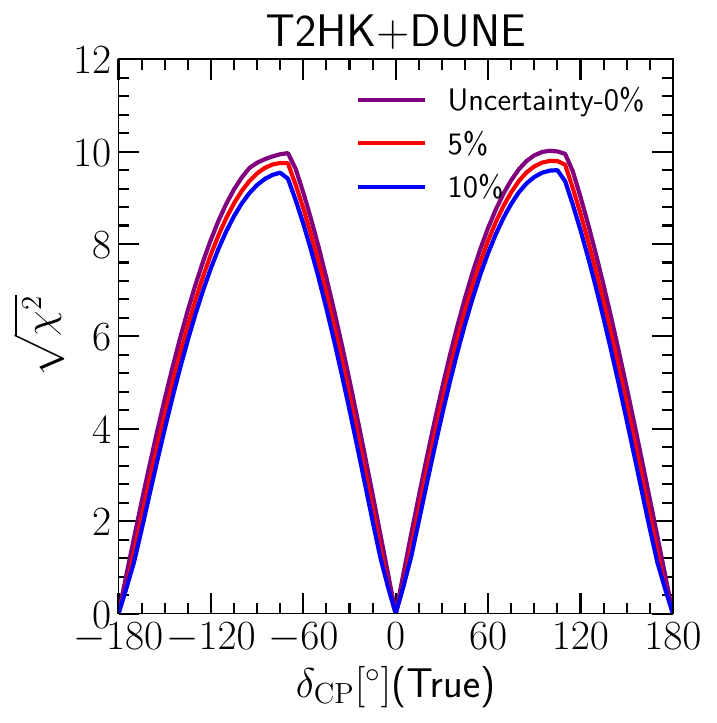}   \\  
\caption{CP violation discovery sensitivity as a function of $\dcp$ (true). Hierarchy unknown. $\theta_{23} = 42^\circ$ is assumed.}
\label{cpv1}
\end{center}
\end{figure}

The significance of rejecting CP invariance is plotted as a function of the true value of $\dcp$ in Fig.\ref{cpv1} in the case of $\theta_{23} = 42^\circ$ for three different reference values (0\%, 5\%, and 10\%) for the uncertainty in the test density variable $\rho$. The top left panel is for T2HK, the top right panel is for DUNE and the bottom right panel is for T2HK+DUNE. For T2HK, sensitivity is good if the true value of $\dcp$ is close to $-90^\circ$,
but for $\sin\dcp\ge 0$ sensitivity is poor because T2HK alone cannot resolve sign degeneracy. However, for DUNE $-$ which can resolve sign degeneracy $-$ sensitivity is good unless $|\sin\dcp|$ is close to 0. Since the matter effect is more important for DUNE, the uncertainty in the density has some effect of sensitivity. The fraction of true $\dcp$ values for which CP violation can be discovered at a $5\sigma$ is approximately 20\%, 35\% and 50\% for T2HK, DUNE and their combination, respectively.

%Fig.3

\begin{figure}[H]
\begin{center}
\includegraphics[scale=0.6]{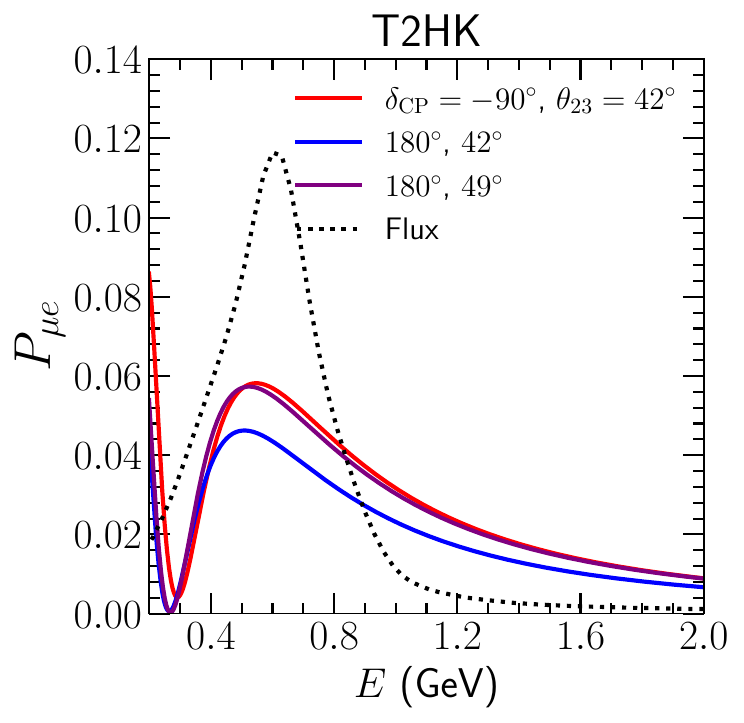}
\includegraphics[scale=0.6]{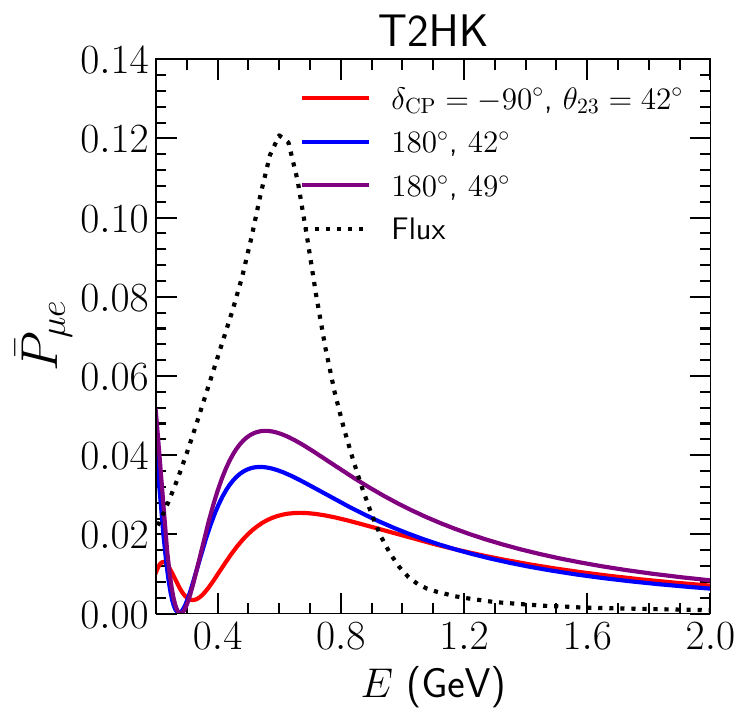} \\     
\includegraphics[scale=0.62]{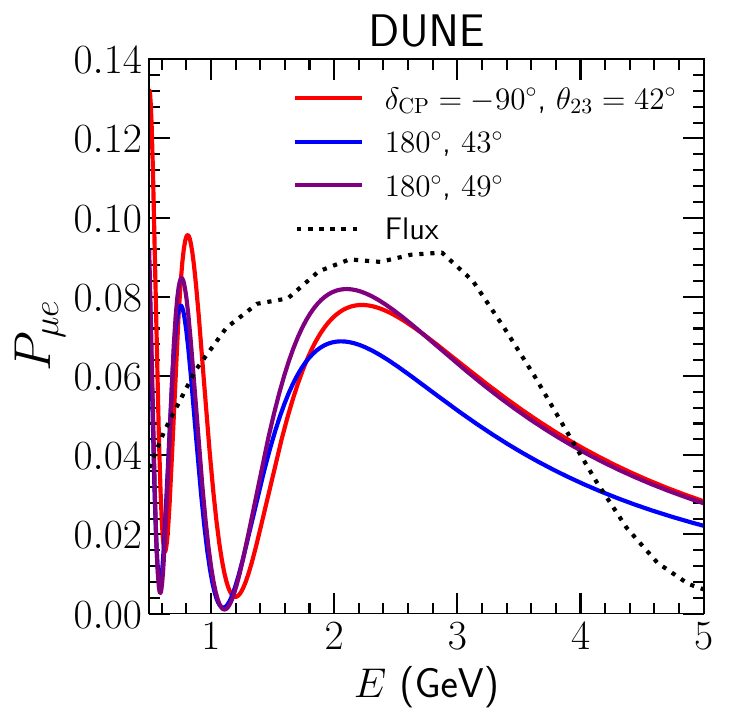}
\includegraphics[scale=0.62]{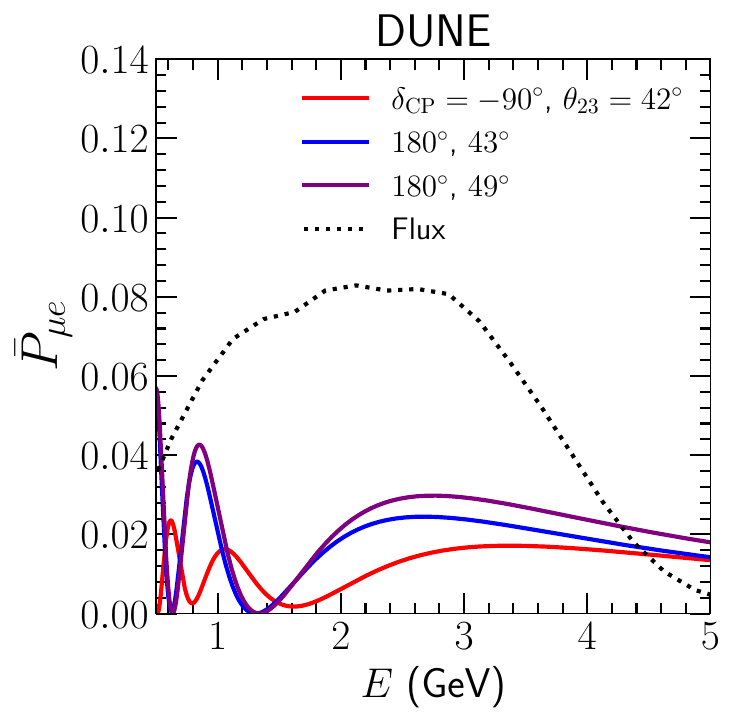}   \\  
\caption{Probability plots to explain CP violation. The appearance probabilities are defined as
$P_{\mu e}\equiv P(\nu_\mu\to\mu_e)$ and $\bar{P}_{\mu e}\equiv P(\bar{\nu}_\mu\to\bar{\mu}_e)$. At $\dcp=-90^\circ$, there is octant degeneracy in  the neutrino  mode; however, the addition of antineutrinos solves the degeneracy.  Flux is also shown in a dotted line. The antineutrino component of DUNE is insufficient to solve the degeneracy. }
\label{cpv2}
\end{center}
\end{figure}

We also observed that there are some strange behaviors for the DUNE plot for $\dcp \sim -90^\circ$ (top right panel of Fig.~\ref{cpv1}). The difference between the two cases $-$ one with a known octant and the other with an unknown octant, shown in the bottom left panel of Fig.~\ref{cpv1} $-$ turns out to be the effect of  some kind of octant degeneracy, where by ``octant degeneracy'' we mean the approximate equality
\begin{eqnarray}
&{\ }& \hspace{-40mm}
P(\nu_\mu\to \nu_e;\theta_{23}=42^\circ,\dcp=-90^\circ)
\simeq
P(\nu_\mu\to \nu_e;\theta_{23}=49^\circ,\dcp=180^\circ)
\label{octant-degeneracy}
\end{eqnarray}
which can be observed in Fig.\ref{cpv2}, where we have plotted the appearance channel probabilities as a function of energy $E$ in GeV. The top row is for T2HK and the bottom row is for DUNE. In each row, the left panel corresponds to neutrinos and the right panel corresponds to antineutrinos. 
%Fig.4
The energy region covered by the flux shows the values of the energies which is relevant for a particular experiment. Note that the approximate equality given by Eq.(\ref{octant-degeneracy}) does not hold for the antineutrino sector, as can be observed also from Fig.\ref{cpv2}. This is a phenomenon different from the original octant degeneracy because in the latter case, the appearance probabilities satisfy the equality for both neutrinos and antineutrinos. We believe that T2HK does not suffer from this octant degeneracy  because the contribution from antineutrinos is large compared with
the one at DUNE; thus, CP invariance is rejected.

The value of $\theta_{23}$ in the wrong octant can be approximately estimated analytically, and it is described in the appendix \ref{appendixa}.

%Fig.5
\begin{figure}[H]
\begin{center}
\includegraphics[scale=0.42]{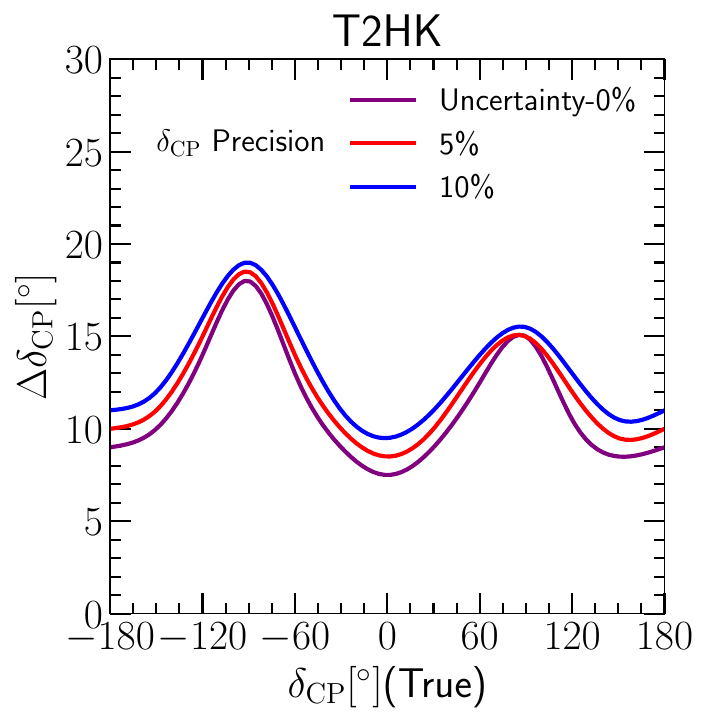}
\includegraphics[scale=0.42]{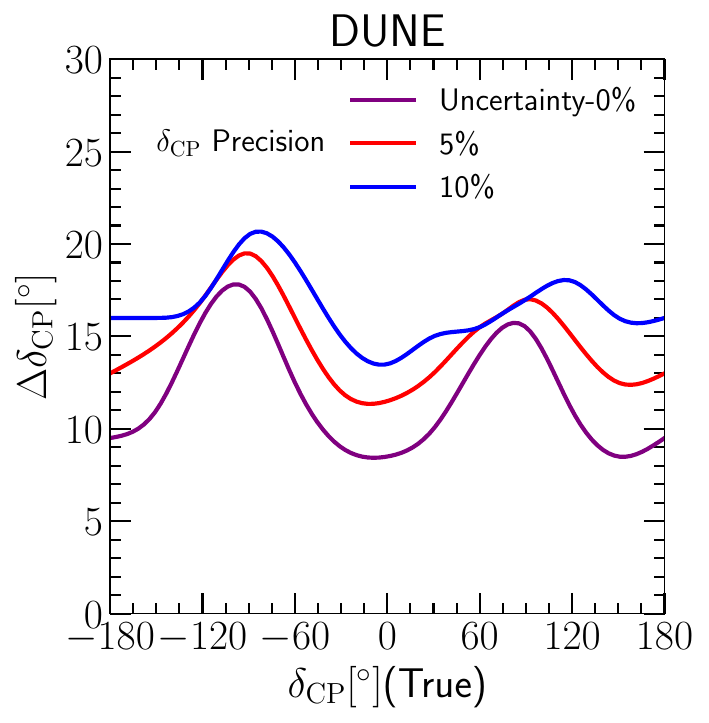}     
\includegraphics[scale=0.42]{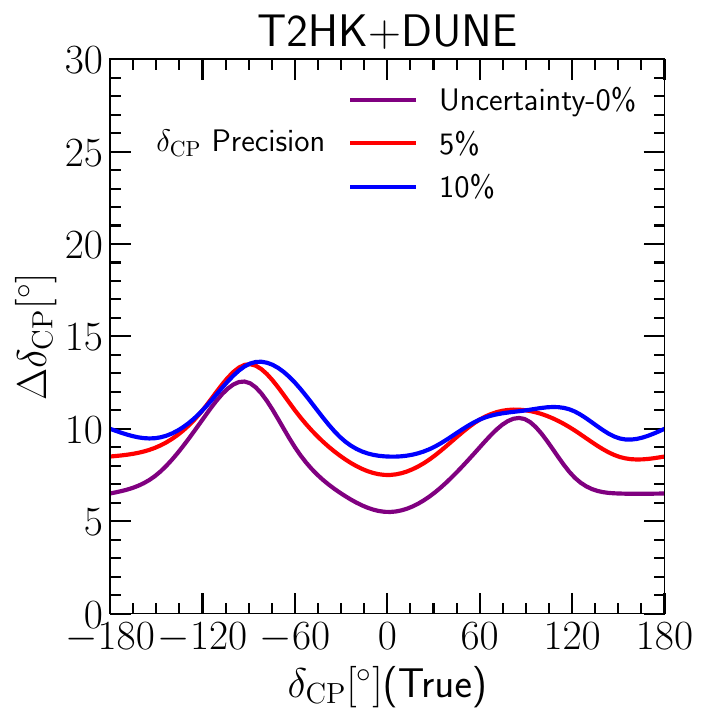} \\
\includegraphics[scale=0.39]{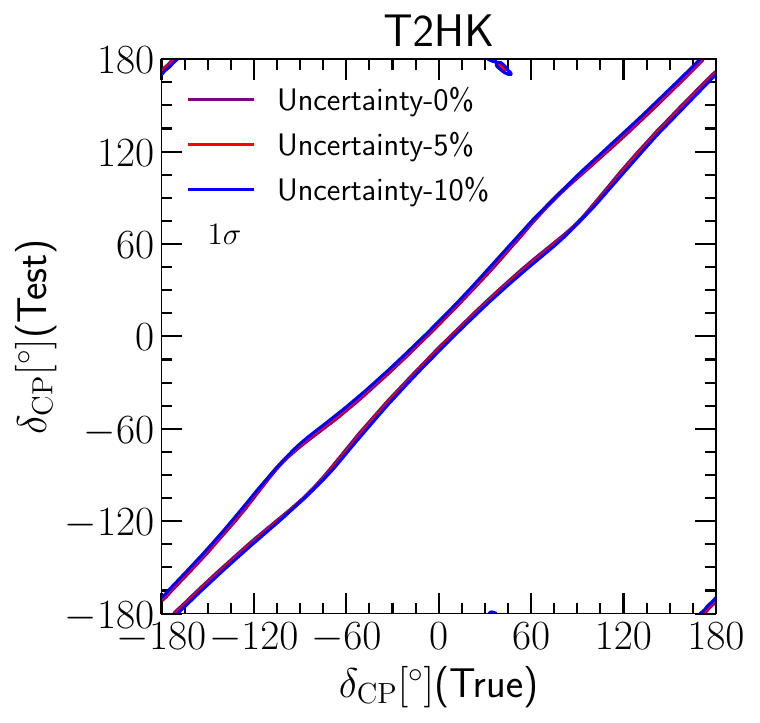}
\includegraphics[scale=0.39]{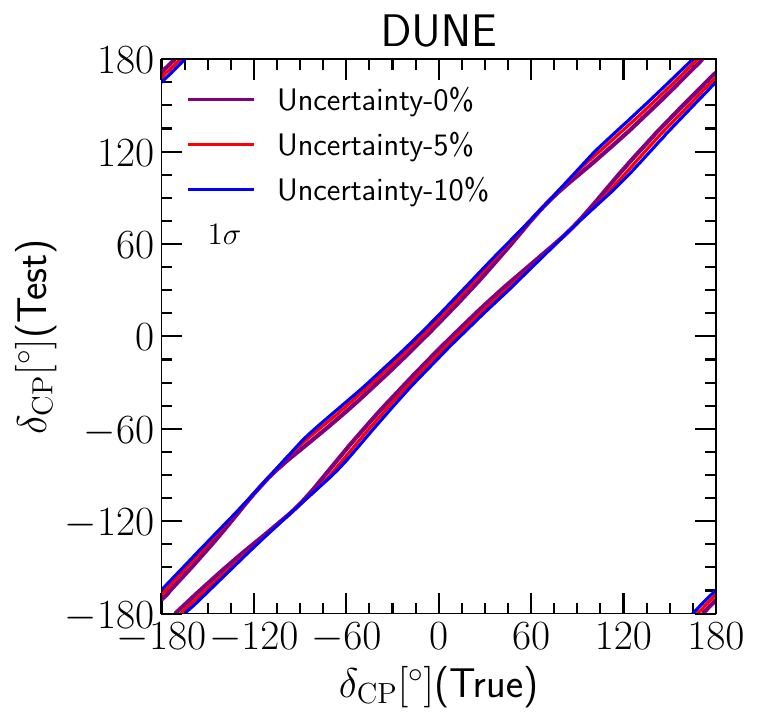}     
\includegraphics[scale=0.39]{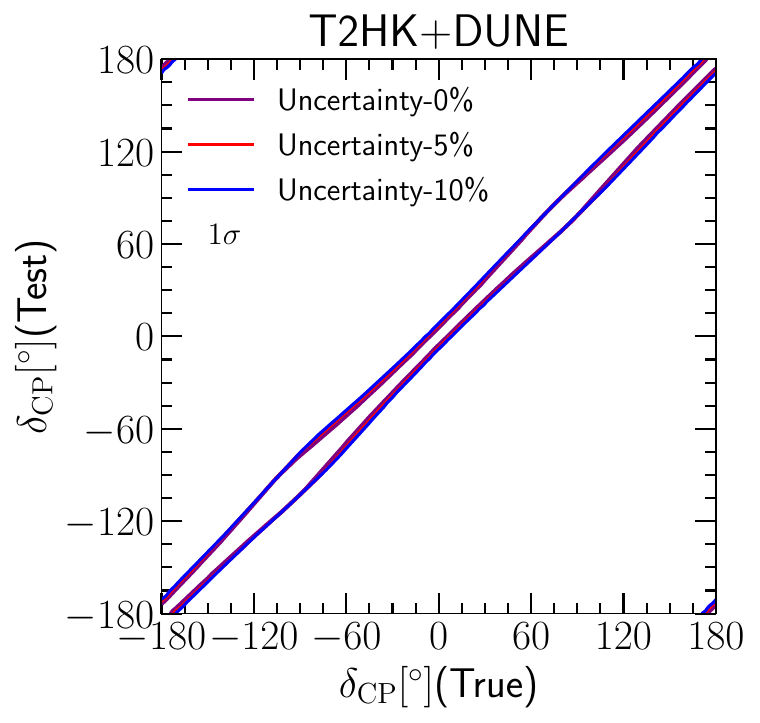}
\caption{$1 \sigma$ CP precision as a function of $\dcp$ (true). Hierarchy unknown. $\theta_{23} = 42^\circ$.}
\label{cpp}
\end{center}
\end{figure}

%\newpage
In Fig.\ref{cpp}, we show the CP precision sensitivity of the experiments that are under consideration. In each row, the left panel is for T2HK, the middle panel is for DUNE and the right panel is for T2HK+DUNE. In the top panel of Fig.\ref{cpp}, we plot the $1 \sigma$ error of the measured value of $\dcp$.  Although the error depends on the true value of $\dcp$, the error of $\dcp$ lies between $7^\circ$ and $20^\circ$ for both T2HK and DUNE, and is between $5^\circ$ and $15^\circ$ for the combined case.  Since DUNE has a longer baseline length, the effect of the uncertainty in the density is larger than that for T2HK, as is naively expected.  From the figures we observe that the uncertainty in the density gives non-negligible contribution
to the precision of $\dcp$, particularly for DUNE.

On the other hand, the correlation of the true and test CP phases is shown in the bottom panel of Fig.\ref{cpp}. Although this figures naively offer the impression that parameter degeneracy (mainly sign degeneracy) is  significantly less serious even for T2HK than Fig.8 in Ref.\!\cite{Fukasawa:2016yue}, in which the analysis was performed with the old design of T2HK with two 0.19 Mton tanks in Japan, sign degeneracy in the bottom panel of Fig.\ref{cpp} is more serious than that in Ref.\!\cite{Fukasawa:2016yue} because significance in Fig.\ref{cpp} is only $1 \sigma$  whereas it is $3 \sigma$ in Fig.8 in Ref.\!\cite{Fukasawa:2016yue}.

%Fig.6
\begin{figure}[H]
\begin{center}
\includegraphics[scale=0.6]{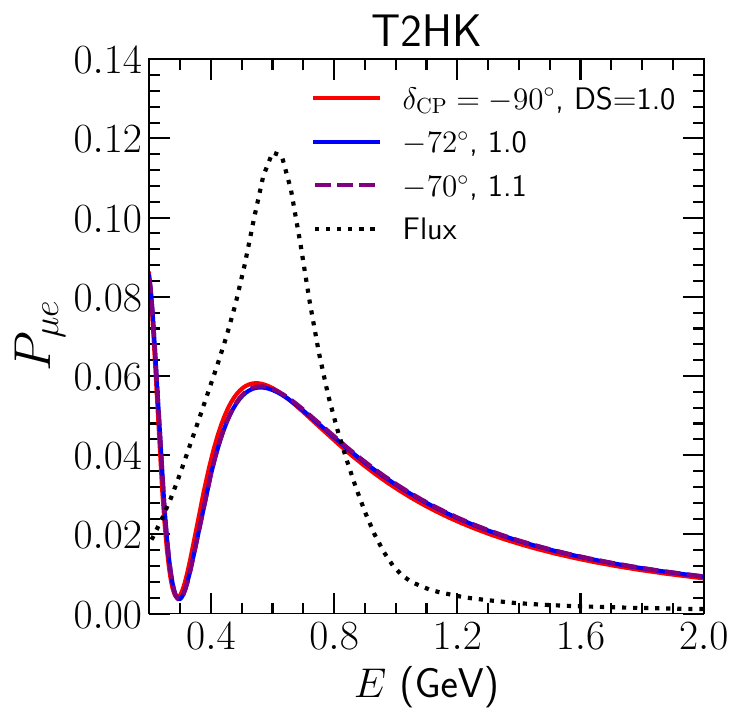}
%\hspace{0.8 in}   
\includegraphics[scale=0.6]{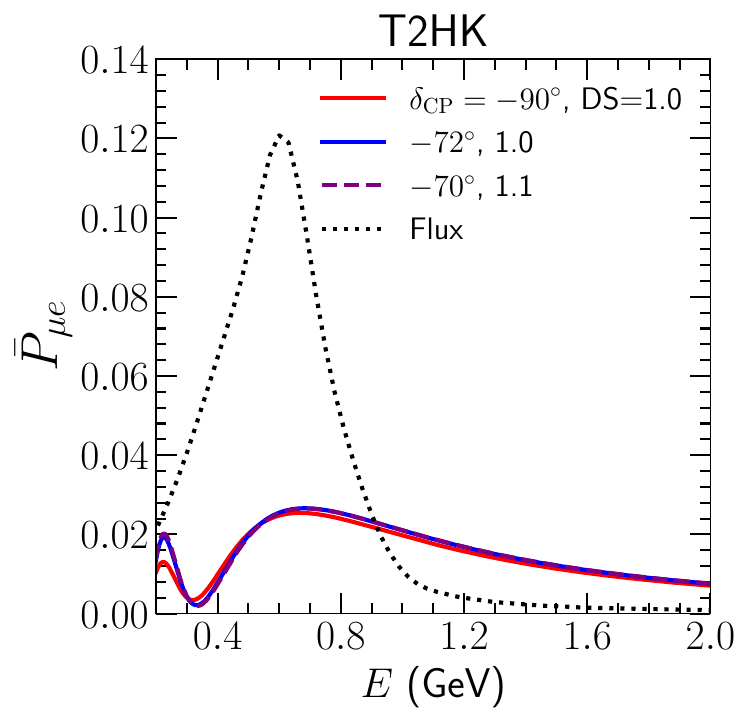} \\     
\includegraphics[scale=0.6]{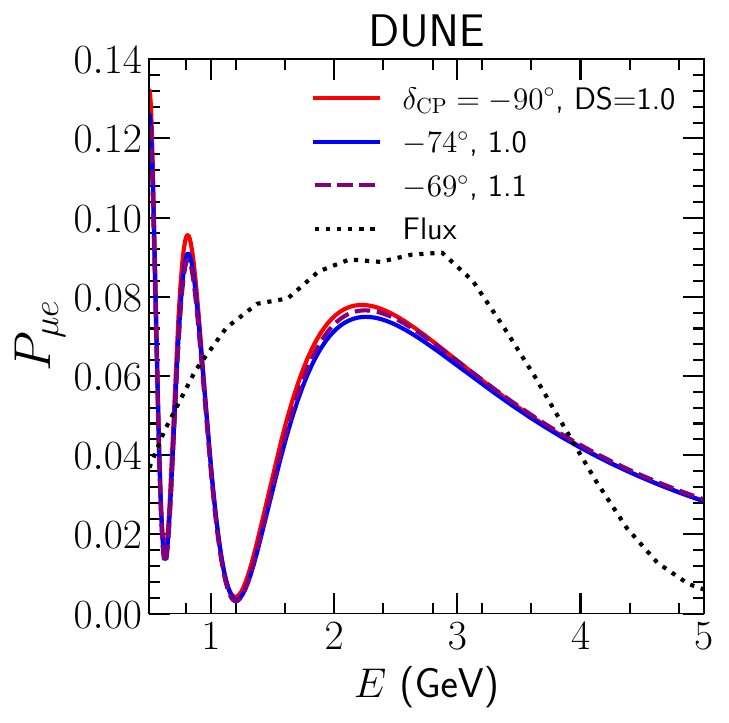}
%\hspace{0.8 in}
\includegraphics[scale=0.6]{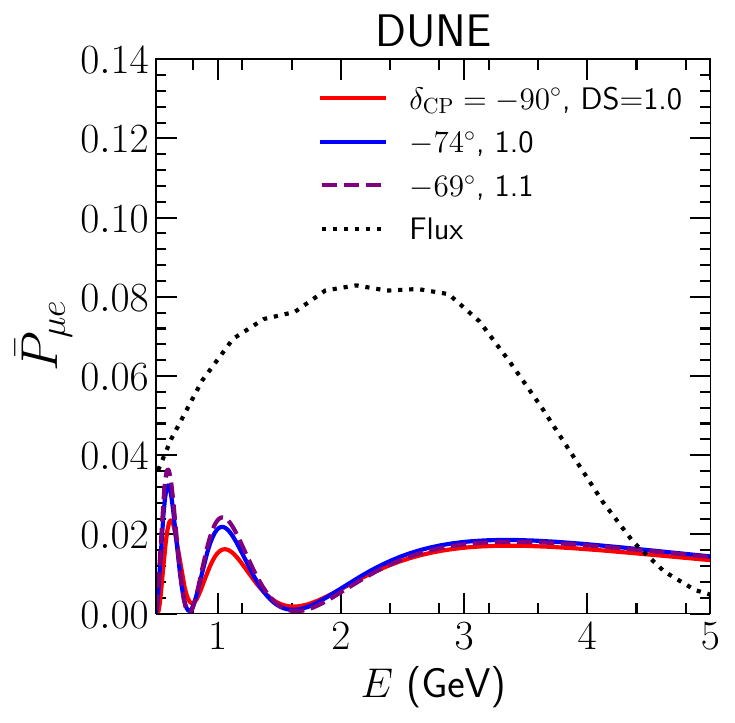}   \\  
\caption{Probability plots to explain CP precision. `DS' corresponds to Density Scaling, which refers to a factor by which density is multiplied. For T2HK, $\dcp = - 72^\circ$ with density $\rho$ is degenerate with ($ - 70^\circ$, $1.1 \rho$). For $\dcp (\rm true) = -90^\circ$ and true density $\rho$, $\dcp (\rm test) = -70^\circ$ is excluded at $1 \sigma$ when the test density is $\rho$\; however, it becomes allowed when test density is $1.1 \rho$. Therefore, the precision deteriorates by 2$^\circ$. Alternatively, the precision deteriorates by 5$^\circ$ for DUNE. Evidently, the amount of
degeneracy is minimal; however, it significantly influences the CP precision.}
\label{cpv3}
\end{center}
\end{figure}

Fig.\ref{cpv3} shows the appearance probabilities $P_{\mu e}\equiv P(\nu_\mu\to\mu_e)$ and $\bar{P}_{\mu e}\equiv P(\bar{\nu}_\mu\to\bar{\mu}_e)$
are plotted as a function of $E$ in GeV  for a few case cases of $\dcp$ and density scaling 'DS' to illustrate how the uncertainty in the Earth's density  can cause an incorrect value for the CP phase. The density scaling refers to a factor by which the the density $\rho$ is multiplied. The top row is for T2HK and the bottom row is for DUNE. In each row, the left panel corresponds to neutrinos and the right panel corresponds to antineutrinos. From the panels we understand that: 
\begin{eqnarray}
&{\ }& \hspace{-40mm}
P(\nu_\mu\to \nu_e;1.0\rho,\dcp=-72^\circ)
\simeq
P(\nu_\mu\to \nu_e;1.1\rho,\dcp=-70^\circ) ~ {\rm for} ~{\rm T2HK}
\label{cpp-degeneracy-t2hk}
\end{eqnarray}
\begin{eqnarray}
&{\ }& \hspace{-40mm}
P(\nu_\mu\to \nu_e;1.0\rho,\dcp=-74^\circ)
\simeq
P(\nu_\mu\to \nu_e;1.1\rho,\dcp=-69^\circ) ~ {\rm for} ~{\rm DUNE}
\label{cpp-degeneracy-dune}
\end{eqnarray}
The above condition is true for both neutrinos and antineutrinos. For this reason, in the case of T2HK, for a true value of $\dcp = -90^\circ$ and true density $\rho$, $\dcp$ (test) $= -72^\circ$ is allowed at $1 \sigma$ when the test density is $\rho$. However, the test value of $\dcp = -70^\circ$ becomes also allowed at $1\sigma$ when the test density is $1.1\rho$, thus deteriorating the CP precision by $2^\circ$. For DUNE, the CP precision deteriorates by $5^\circ$, when the test density becomes $1.1\rho$.  Therefore we understand that 
a small error in the density can result in an incorrect value of $\dcp$ when we consider both $P_{\mu e}$ and $\bar{P}_{\mu e}$.

%Fig.7
\begin{figure}[H]
\begin{center}
\includegraphics[scale=0.44]{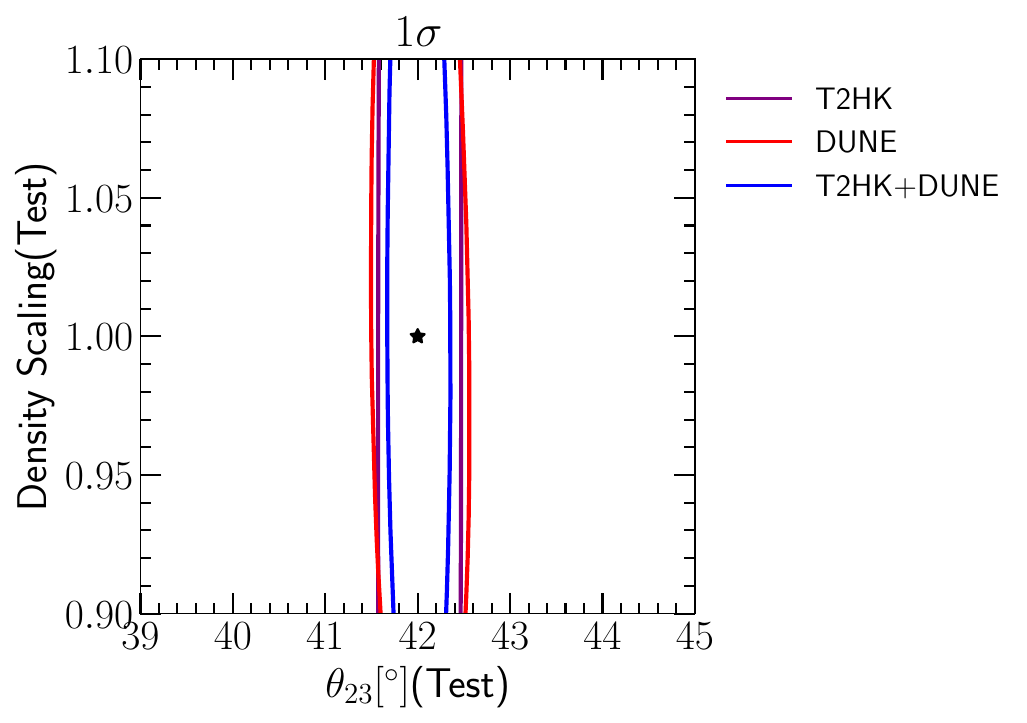}
\includegraphics[scale=0.44]{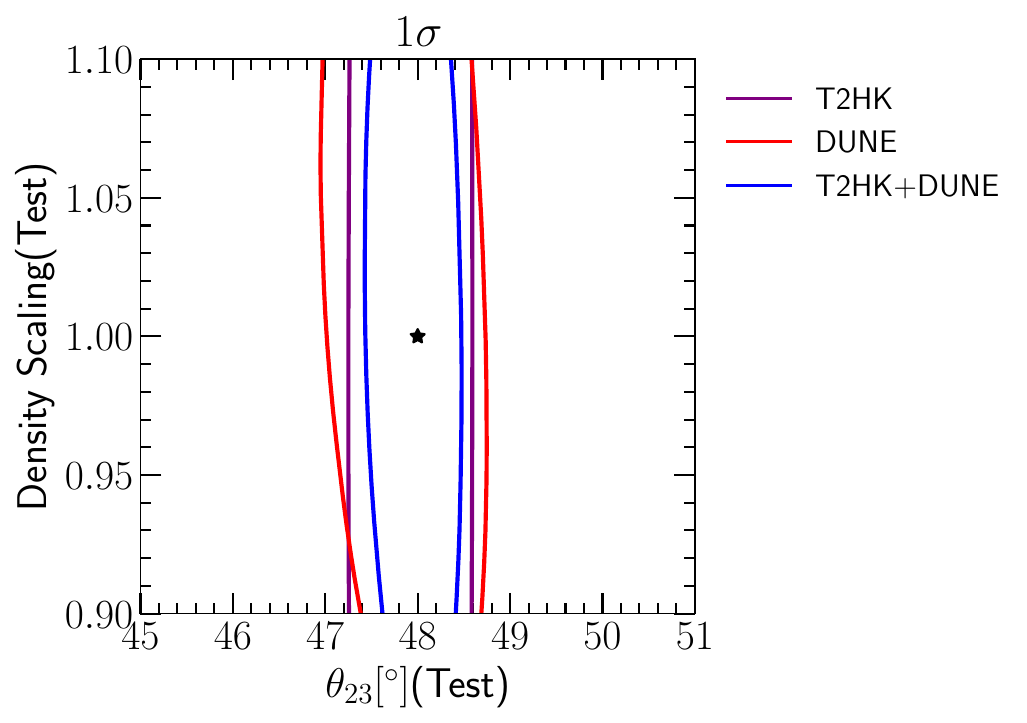} \\
\includegraphics[scale=0.44]{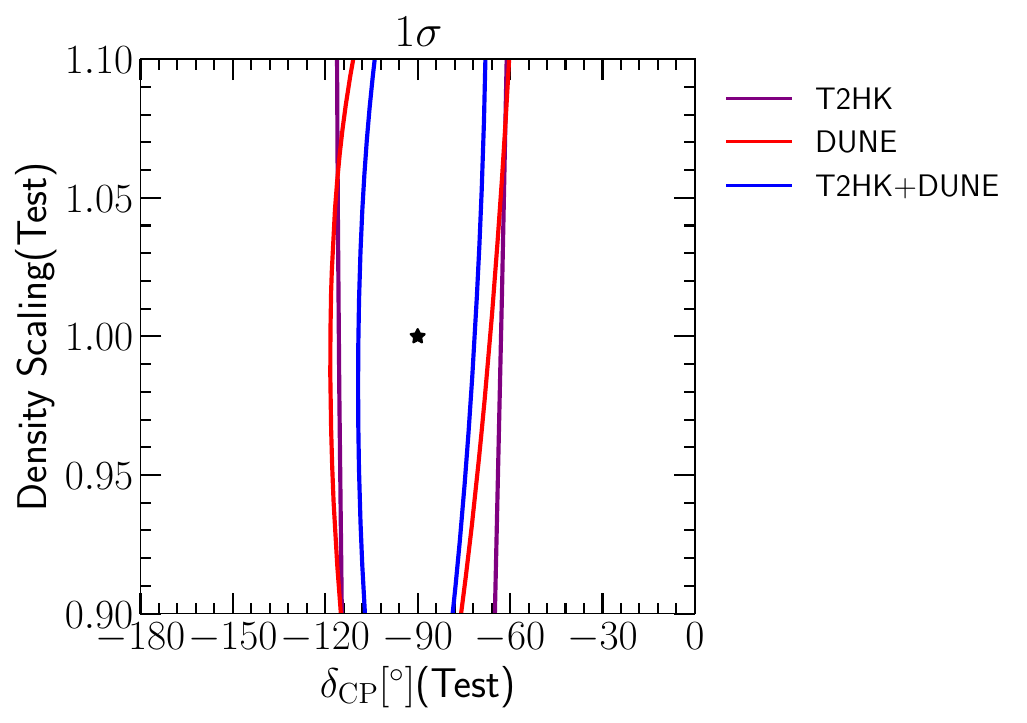}   
\includegraphics[scale=0.44]{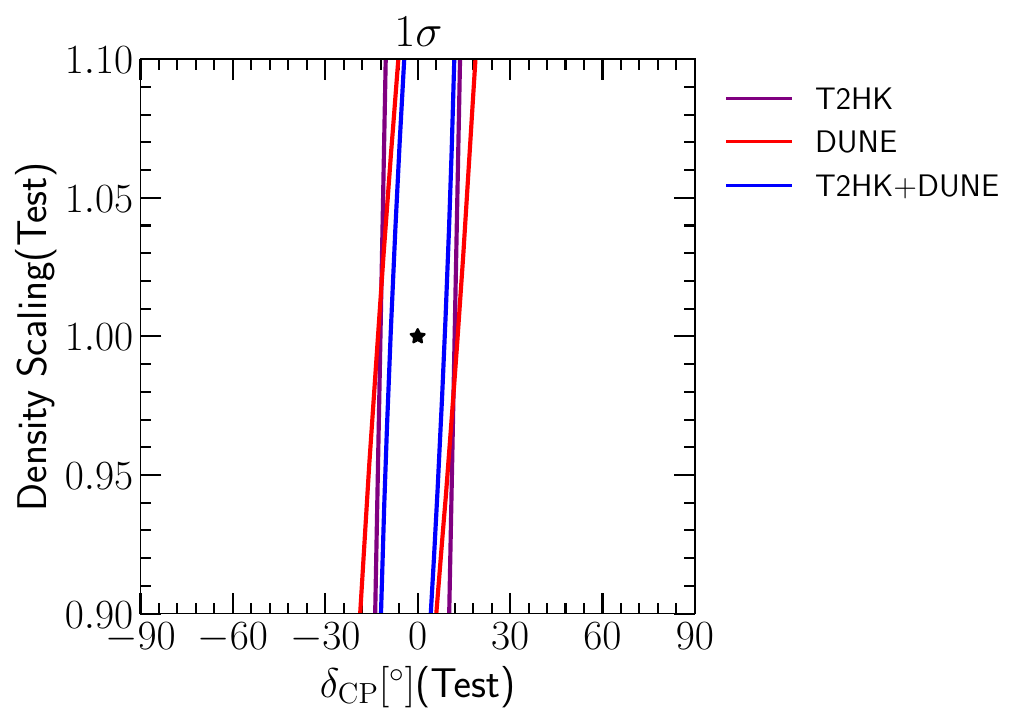}   
\includegraphics[scale=0.44]{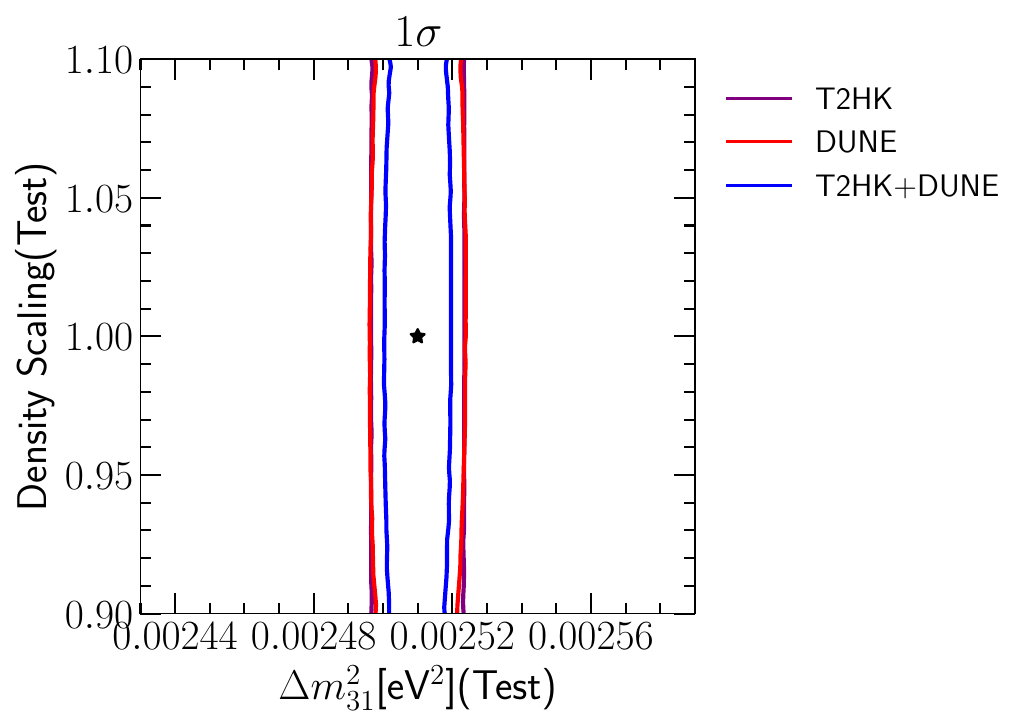} 
\caption{precision of $\dcp$ - density ($\theta_{23} =
42^\circ$), $\theta_{23}$ - density ($\dcp = -90^\circ$)
and $\Delta m^2_{31}$ - density ($\dcp = -90^\circ$,
$\theta_{23} = 42^\circ$). Hierarchy known.}
\label{octb}
\end{center}
\end{figure}

In Fig.\ref{octb}, we show the dependence of the test density variable on the $1 \sigma$ allowed region for the oscillation parameters $\theta_{23}$ (top row), $\dcp$ (middle row) and $\Delta m^2_{31}$ (bottom row). For $\theta_{23}$ we have considered two true values of $42^\circ$ (top left panel) and $48^\circ$ (top right panel), for $\dcp$ we have considered two true vales of $-90^\circ$  (middle left panel) and $0^\circ$ (middle right panel) and for $\Delta m^2_{31}$ we have considered the true value as $2.510 \times 10^{-3}$ eV$^2$. The density scaling is 1.0 in the true spectrum for all these panels. In calculation of this region, the test density variable is not varied but fixed and plotted in the y axis. Since both T2HK and DUNE experiments
have a baseline length shorter than 4000 km for which the contribution of the matter effect to the oscillation probability would be of order 1, dependence of the allowed region on the density scaling factor is small. From the panels we see that for $\theta_{23}$ and $\Delta m^2_{31}$ the curves are almost parallel to the $y$ axis. This implies the fact that the measurement of these parameters are not much affected due to the uncertainty in the density. This we will further see in Fig.~\ref{prec}. However, for $\dcp$, we notice that the curves are slightly tilted, implying the fact that uncertainty in the density can alter the precision measurement of $\dcp$ which we have seen in Fig.~\ref{cpp}.

%\newpage
\subsection{Precision in measurements of $\Delta m^2_{31}$ and $\theta_{23}$}
%Fig.8
\begin{figure}[H]
\begin{center}
\includegraphics[scale=0.39]{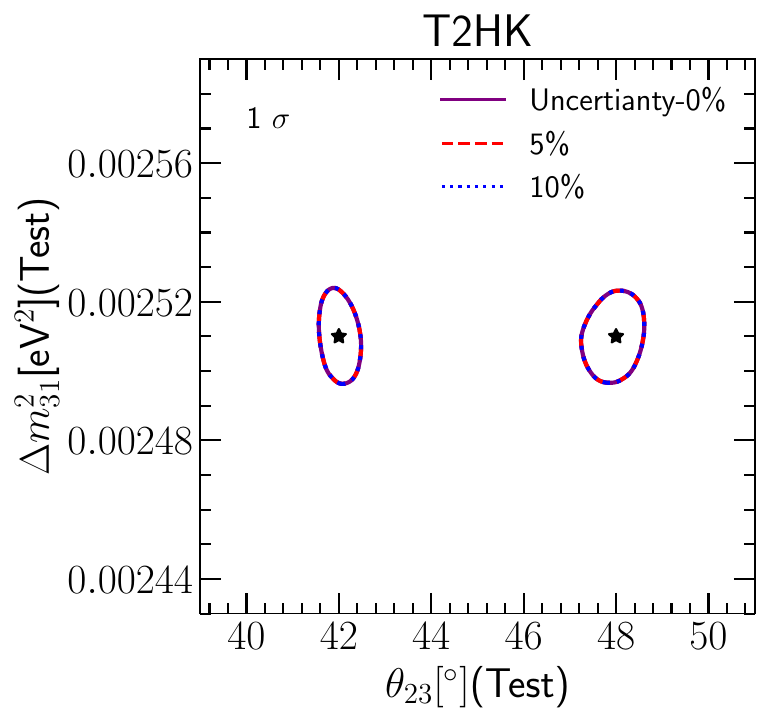}
\includegraphics[scale=0.39]{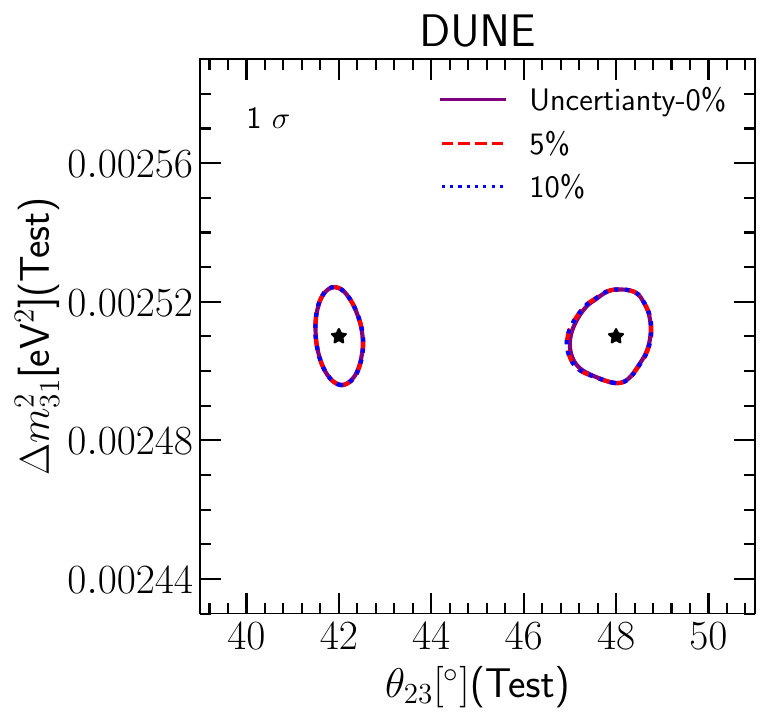}     
\includegraphics[scale=0.39]{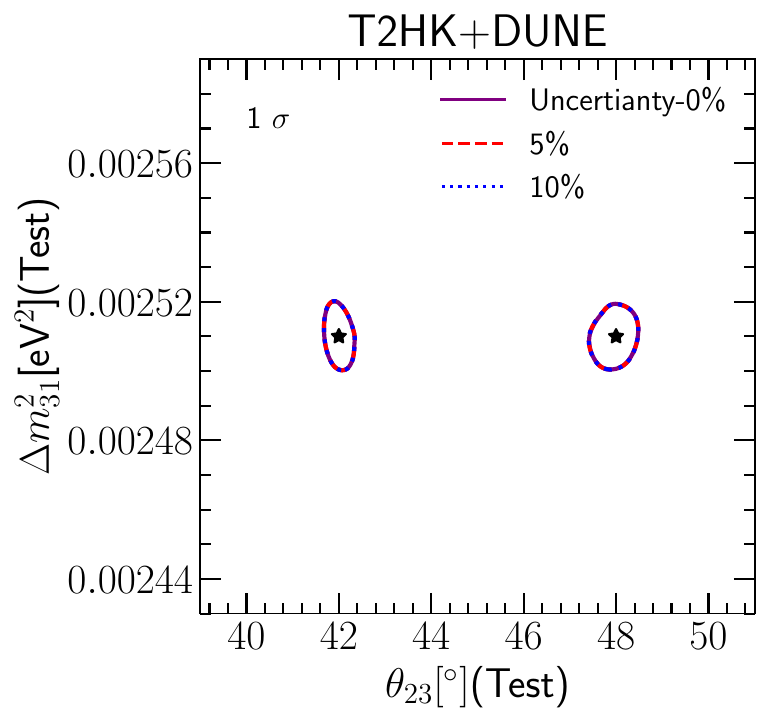}
\caption{$1 \sigma$ precision of $\theta_{23}$ and $\Delta
m^2_{31}$. Mass ordering is assumed to be known,
and $\dcp = -90^\circ$ is assumed.}
\label{prec}
\end{center}
\end{figure}

In Fig.\ref{prec}, we plot the $1 \sigma$ allowed region in the ($\theta_{23}$, $\Delta m^2_{31}$)-plane, where normal mass ordering and $\dcp = -90^\circ$ is assumed. The true values of ($\theta_{23}$, $\Delta m^2_{31}$) are marked by a star. The left panel is for T2HK, the middle panel is for DUNE and the right panel is for T2HK+DUNE. From Fig.\ref{prec}, we obtain the $1 \sigma$ allowed region around the true values of $\theta_{23}$ and
$\Delta m^2_{31}$ and they are given in Table.\ref{table3}. The precision of T2HK, DUNE and their combination in measurements of $\Delta m^2_{31}$ and $\theta_{23}$ is better than the present one which is obtained of the global fits.\cite{Gonzalez-Garcia:2021dve,Capozzi:2021fjo,deSalas:2020pgw}
The oscillation parameters $\theta_{23}$ and $\Delta m^2_{31}$ are determined primarily from the disappearance channel whose oscillation
probability has small dependence on $\dcp$. Therefore the uncertainty in the density has little effect on the allowed region in the ($\theta_{23}$, $\Delta m^2_{31}$)-plane at $1 \sigma$.

\begin{table} [H]
\centering
\begin{tabular}{|c|c|c|c|} \hline
Parameter   &  T2HK &  DUNE & {\small T2HK+DUNE}\\          
\hline
$\theta_{23}$ (LO)
& $41.5^\circ$ - $42.5^\circ$
& $41.5^\circ$ - $42.5^\circ$
& $41.7^\circ$ - $42.4^\circ$
\\ 
$\theta_{23}$ (HO)
& $47.2^\circ$ - $48.6^\circ$
& $46.9^\circ$ - $48.8^\circ$
& $47.4^\circ$ - $48.5^\circ$
\\ 
$\Delta m^2_{31}~[10 ^{-3}\mbox{\rm  eV}^2]$
& 2.496  - 2.523
& 2.496  - 2.525
& 2.500  - 2.520
\\
\hline
\end{tabular}
\caption{$1 \sigma$ allowed region for $\theta_{23}$ and $\Delta m^2_{31}$.  LO (HO) represents the case in which the true value for $\theta_{23}$ is $42^\circ$ ($48^\circ$).  Normal mass ordering is assumed as the true mass ordering.}
\label{table3}
\end{table}

%\newpage
\section{Conclusion}
\label{conclusion}

Here, we studied the sensitivity of T2HK, DUNE and their combination to mass ordering, octant to which $\theta_{23}$ belongs, and $\dcp$ and precision measurement of ($\theta_{23}$, $\Delta m^2_{31}$) by  considering the uncertainty in the Earth's density. The main results of our analysis are  presented in Tables \ref{table4}.

\begin{table}[H]
\begin{center}
%\hspace{-5mm}
\begin{tabular}{|c|c|c|c|c|}
\hline
Sensitivity   & Max $|\Delta\rho/\rho|$& T2HK &  DUNE & {\small T2HK+DUNE}\\          
\hline
Mass ordering& 0\% & 1.03 &7.8  & 11.7    \\
significance  & 5\% & 1.00 &7.5  & 11.4    \\
$[\sigma]$& 10\% & 0.96 &7.2  & 11.1    \\
\hline
Octant& 0\%&   $42.343^\circ - 48.674^\circ$   &  $42.21^\circ - 49.03^\circ$  & $42.96^\circ - 48.00^\circ$   \\
at      & 5\%&   $42.340^\circ - 48.676^\circ$   &  $42.16^\circ - 49.11^\circ$  & $42.93^\circ - 48.03^\circ$   \\
$5 \sigma$& 10\%&   $42.338^\circ - 48.678^\circ$   &  $42.12^\circ - 49.18^\circ$  & $42.93^\circ - 48.06^\circ$   \\
\hline
CP violation& 0\%&    21.4  &  37.4  & 54.9    \\
fraction  & 5\%&    20.8  &  34.2  & 52.7    \\
$[$\%$]$& 10\%&    19.8  &  32.0  & 50.2    \\
\hline
CP Precision & 0\%&    $18^\circ$  &  $16^\circ$  & $12^\circ$    \\
at  & 5\%&    $19^\circ$ &  $19^\circ$  & $13^\circ$    \\
$1\sigma$& 10\%&    $20^\circ$  &  $21^\circ$  & $13^\circ$    \\
\hline

\hline
\end{tabular}
\end{center}
\caption{The sensitivity for  mass ordering, the $\theta_{23}$ octant and CP for conservative values of the true parameter for different choices of the
uncertainty in the density.  For the mass ordering, we provide the values of $\sqrt{\chi^2}$.  Additionally, for the octant $-$ assuming that the true value for ($\theta_{23}$, $\dcp$) is ($45^\circ$, $-90^\circ$) $-$ we provide the region for  which $\theta_{23}$  is permitted at $5 \sigma$.  For CP violation  (CPV), the fraction of the $\dcp$ region for which CPV can be discovered at $\chi^2=25$ is given for $\theta_{23}=42^\circ$. For the CP precision, the $1 \sigma$ error for $\dcp = -90^\circ$ is given.}
\label{table4}
\end{table}

Notably, the T2HK experiment does not have good sensitivity to mass ordering; i.e., the significance  ranges between $1 \sigma$ and $5 \sigma$. However, it can resolve octant degeneracy at $5 \sigma$ if $\theta_{23}$ lies outside the region $42.3^\circ\lesssim\theta_{23}\lesssim 48.7^\circ$.
Since T2HK does not have good sensitivity to mass ordering, the fraction of the $\dcp$ region for which CPV can be discovered at $5 \sigma$ is limited.  Conversely, DUNE has good sensitivity to mass ordering $-$ with a significance ranging between $8 \sigma$ and $18 \sigma$ $-$ and it can resolve the octant degeneracy at the $5 \sigma$ level if $\theta_{23}$ lies outside the region $42.2^\circ\lesssim\theta_{23}\lesssim 49.1^\circ$.  The fraction of the $\dcp$ region for which DUNE can discover CPV at $5 \sigma$ level is larger than that for T2HK because DUNE has good sensitivity to mass ordering. Regarding the CP precision, the $1 \sigma$ error corresponding to a measurement of $\dcp$ ranges from $7^\circ$ to $20^\circ$ for both T2HK and DUNE.  Because of the longer baseline DUNE suffers from the uncertainty in the Earth's density  more than T2HK does, and the
sensitivity of DUNE depends more on the uncertainty in the density.  Furthermore, when T2HK and DUNE are combined, we obtain the best sensitivity  for mass ordering, the $\theta_{23}$ octant, a large fraction of the $\dcp$ region for which we can discover CPV at $5 \sigma$, and also the CP precision. The combination of T2HK and DUNE will provide a better precision for the oscillation parameters $\theta_{23}$ and $\Delta m^2_{31}$
than what is currently provided by the global fits.

Moreover, the effects of the uncertainty in the Earth's density  on the measurements of the oscillation parameters for T2HK and DUNE are small compared with that expected at a neutrino factory.  However, with an uncertainty of 5\% we observed that the CP precision of DUNE has a non-negligible impact. Therefore, when the precision of the CP phase $\dcp$ becomes important in the future, the effect of the uncertainty in the density should be considered properly.

In this study, we did not include the atmospheric neutrino measurement at Hyper-kamiokande.  If information of the atmospheric neutrino of Hyper-kamiokande is included, then the sensitivity of the whole Hyper-kamiokande project, which includes the T2HK experiment and the atmospheric neutrino measurement, to mass ordering will be improved, as shown in Refs.\!\cite{Hyper-KamiokandeWorkingGroup:2014czz,Fukasawa:2016yue,Hyper-Kamiokande:2016srs}.
A treatment of the uncertainty in the density for the analysis of atmospheric neutrinos is complicated and it is a subject in the future work.

\section*{Acknowledgments}
This research was partly supported by a Grant-in-Aid for Scientific Research of the Ministry of Education, Science and Culture, under Grants No. 18H05543 and No. 21K03578. MG acknowledges Ramanujan Fellowship of SERB, Govt. of India, through grant no: RJF/2020/000082.

\newpage
%\vglue 10mm
\noindent
{\Large\bf Appendix}

\appendix

\section{Octant degeneracy}
\label{appendixa}

\subsection{Appearance probability}
Propagation of neutrinos in  the matter is described by
\begin{eqnarray}
&{\ }& \hspace{-80mm}
\displaystyle i {d \over dt} \left( \begin{array}{c} \nu_e(t) \\
\nu_\mu(t)\\\nu_\tau(t)
\end{array} \right)
=M~\left( \begin{array}{c} \nu_e(t) \\
\nu_\mu(t)\\\nu_\tau(t)
\end{array} \right)\,,
\label{sch3}
\end{eqnarray}
where the Hamiltonian $M$ can be diagonalized by a unitary matrix $\tilde U$ as
\begin{eqnarray}
&{\ }&\hspace*{-66mm}M = U {\cal E} U^{-1} + {\cal A}  
=\tilde U \tilde{\cal E} \tilde U^{-1}\,,
\nonumber\\
&{\ }&\hspace*{-66mm}
{\cal E} = \mbox{\rm diag}(0,\Delta E_{21},\Delta E_{31})\,,
\nonumber\\
&{\ }&\hspace*{-66mm}
\Delta E_{jk}\equiv\frac{\Delta m^2_{jk}}{2E}\,,
\nonumber\\
&{\ }&\hspace*{-66mm}
{\cal A} = \mbox{\rm diag}(A,0,0)\,,
\nonumber\\
&{\ }&\hspace*{-66mm}
A\equiv\sqrt{2}G_FN_e\,,
\nonumber\\
&{\ }&\hspace*{-66mm}
\tilde{\cal E} = \mbox{\rm diag}
(\tilde{E}_1,\tilde{E}_2,\tilde{E}_3)\,.
\nonumber
\end{eqnarray}
$G_F$ and $N_e$ represent the Fermi coupling constant and the density of electrons, respectively, and $\tilde{E}_j~(j=1,2,3)$ is the energy eigenvalue of the Hamiltonian $M$. If $N_e$ is constant, then Eq.(\ref{sch3}) can be solved as
\begin{eqnarray}
&{\ }&\hspace*{-56mm} 
\nu_\beta (t) = \sum_\alpha\left[\tilde U \exp\left(
-i\tilde{\cal E}t \right)\tilde U^{-1}
\right]_{\beta\alpha}
\nu_\alpha (0)\,.
\nonumber
\nonumber
\end{eqnarray}
Therefore, starting with the initial flavor eigenstate $\nu_\alpha$, the probability amplitude to obtain the flavor eigenstate $\nu_\beta$ after running for a distance of $L$ is given by
\begin{eqnarray}
&{\ }&\hspace*{-46mm} 
A(\nu_\alpha\to\nu_\beta)=\left[\tilde U \exp\left(
-i\tilde{\cal E}L \right)\tilde U^{-1}\right]_{\beta\alpha}
\nonumber\\
&{\ }&\hspace*{-22mm} 
=\sum_{j=1}^3 \left(\tilde U\right)_{\beta j}
\exp\left(-i\tilde{E}_jL \right)
\left(\tilde U^{-1}\right)_{j \alpha}
\nonumber\\
&{\ }&\hspace*{-22mm} 
=\sum_{j=1}^3 \tilde U_{\beta j}
\tilde U_{\alpha j}^\ast
\exp\left(-i\tilde{E}_jL \right)
\nonumber\,.
\end{eqnarray}
As in vacuum, the appearance probability $P_{\mu e}\equiv P(\nu_\mu\to\nu_e)$ is given by the absolute square of the probability amplitude $A(\nu_\mu\to\nu_e)$ and it can be calculated as follows:
\begin{eqnarray}
&{\ }&\hspace*{-16mm}P_{\mu e} = 
\left|
\sum_{j=1}^3 \tilde{U}_{ej}\tilde{U}_{\mu j}^\ast e^{-i\tilde E_j L}
\right|^2
\nonumber\\
&{\ }&\hspace*{-9mm}= \left|e^{-i\tilde E_1 L}
\sum_{j=1}^3 \tilde{U}_{ej}\tilde{U}_{\mu j}^\ast e^{-i\Delta \tilde E_{j1} L}
\right|^2
\nonumber\\
&{\ }&\hspace*{-9mm}= \left|
\sum_{j=1}^3 \tilde{U}_{ej}\tilde{U}_{\mu j}^\ast\left(e^{-i\Delta \tilde E_{j1} L}-1\right)
\right|^2
\nonumber\\
&{\ }&\hspace*{-9mm}
\quad\quad(~\mbox{\rm unitarity~identity}
~\sum_{j=1}^3 \tilde{U}_{\mu j}\tilde{U}_{ej}^\ast=0
~\mbox{\rm was~subtracted})
\nonumber\\
&{\ }&\hspace*{-9mm}= \left|
(-2i)\sum_{j=1}^3
e^{-i\Delta \tilde E_{j1} L/2}
\tilde{U}_{e3}\tilde{U}_{\mu 3}^\ast 
\sin\left(\frac{\Delta \tilde E_{j1} L}{2}\right)
\right|^2
\nonumber\\
&{\ }&\hspace*{-9mm}= 4\left|
e^{-i\Delta \tilde E_{31} L/2}
\tilde{U}_{e3}\tilde{U}_{\mu 3}^\ast 
\sin\left(\frac{\Delta \tilde E_{31} L}{2}\right)
+2ie^{-i\Delta \tilde E_{21} L/2}
\tilde{U}_{e2}\tilde{U}_{\mu 2}^\ast 
\sin\left(\frac{\Delta \tilde E_{21} L}{2}\right)
\right|^2
\nonumber\\
&{\ }&\hspace*{-9mm}=4 \left|
\tilde{U}_{e3}\tilde{U}_{\mu 3}^\ast 
\sin\left(\frac{\Delta \tilde E_{31} L}{2}\right)
+e^{i\Delta \tilde E_{32} L/2}
\tilde{U}_{e2}\tilde{U}_{\mu 2}^\ast 
\sin\left(\frac{\Delta \tilde E_{21} L}{2}\right)
\right|^2\,,
\nonumber
\end{eqnarray}
where
\begin{eqnarray}
&{\ }&\hspace{-100mm}
\Delta\tilde{E}_{jk}\equiv \tilde{E}_j - \tilde{E}_k
\nonumber
\end{eqnarray}
and the result so far is exact if $A$ is constant.

In the case of atmospheric $\nu$ or accelerator $\nu$, we have $|\Delta E_{31}|\simeq A \gg |\Delta E_{21}|$, so we keep $\Delta E_{31}$ and $A$ and treat  $\Delta E_{21}$ as perturbation, keeping only the first order terms in $\Delta E_{21}$.

\noindent
If $|\Delta E_{31}|\gg A$, then to
zero-th order in $\Delta E_{21}/\Delta E_{31}$, we have
\begin{eqnarray}
&{\ }&\hspace{-40mm}
\tilde{E}_1^{(0)} = \Lambda_- \equiv
\frac{1}{2}\left(
\Delta E_{31}+A-\Delta \tilde{E}_{31}\right)
\label{lambdan}\\
&{\ }&\hspace{-40mm}
\tilde{E}_2^{(0)} = 0
\nonumber\\
&{\ }&\hspace{-40mm}
\tilde{E}_3^{(0)} = \Lambda_+ \equiv
\frac{1}{2}\left(
\Delta E_{31}+A+\Delta \tilde{E}_{31}\right)
\label{lambdap}\\
&{\ }&\hspace{-40mm}
\Delta \tilde{E}_{31}^{(0)}
=\sqrt{(\Delta {E}_{31}\cos2\theta_{13}-A)^2
+(\Delta {E}_{31}\sin2\theta_{13})^2
}
\label{deltatildee31}\\
&{\ }&\hspace{-40mm}
\Delta \tilde{E}_{32}^{(0)}
= \tilde{E}_3^{(0)} = \Lambda_+
\nonumber\\
&{\ }&\hspace{-40mm}
\Delta \tilde{E}_{21}^{(0)}
= -\tilde{E}_1^{(0)} = - \Lambda_-\,.
\nonumber
\end{eqnarray}
From the method by  Kimura, Takamura and Yokomakura\cite{Kimura:2002wd}, we have
\begin{eqnarray}
\left(\begin{array}{c}
\tilde{U}_{e1}\tilde{U}_{\mu 1}^\ast \cr\cr
\tilde{U}_{e2}\tilde{U}_{\mu 2}^\ast \cr\cr
\tilde{U}_{e3}\tilde{U}_{\mu 3}^\ast 
\end{array}\right)
=\left(\begin{array}{ccc}
\displaystyle
\frac{{\ }1}{\Delta \tilde{E}_{21} \Delta \tilde{E}_{31}}
(\tilde{E}_2\tilde{E}_3, & -(\tilde{E}_2+\tilde{E}_3),&
1)\cr
\displaystyle
\frac{-1}{\Delta \tilde{E}_{21} \Delta \tilde{E}_{32}}
(\tilde{E}_3\tilde{E}_1, & -(\tilde{E}_3+\tilde{E}_1),&
1)\cr
\displaystyle
\frac{{\ }1}{\Delta \tilde{E}_{31} \Delta \tilde{E}_{32}}
(\tilde{E}_1\tilde{E}_2, & -(\tilde{E}_1+\tilde{E}_2),&
1)\cr
\end{array}\right)
\left(\begin{array}{r}
\delta_{e\mu}\cr\cr
\left[U{\cal E}U^{-1}+{\cal A}\right]_{e\mu}\cr\cr
\left[\left(U{\cal E}U^{-1}+{\cal A}\right)^2\right]_{e\mu}
\end{array}\right)\,,
\nonumber
\end{eqnarray}
and
$\tilde{U}_{e2}\tilde{U}_{\mu 2}^\ast$ and $\tilde{U}_{e3}\tilde{U}_{\mu 3}^\ast$
can be approximated to zero-th order in $\Delta E_{21}/\Delta E_{31}$ as
\begin{eqnarray}
&{\ }&\hspace{-20mm}
2\tilde{U}_{e3}\tilde{U}_{\mu 3}^\ast\simeq
2\frac{\Delta E_{31}}
{\Delta \tilde{E}_{31}} U_{e3}U_{\mu 3}^\ast
= 2\frac{\Delta E_{31}}
{\Delta \tilde{E}_{31}}\, e^{-i\dcp} s_{13} c_{13} s_{23}
= \frac{\Delta E_{31}}
{\Delta \tilde{E}_{31}}\, e^{-i\dcp} s_{23} \sin2\theta_{13}
\nonumber\\
%\end{eqnarray}
%\begin{eqnarray}
&{\ }&\hspace{-20mm}
2\tilde{U}_{e2}\tilde{U}_{\mu 2}^\ast \simeq 
-2\frac{\Delta E_{21}}{Ac_{13}^2}U_{e2}U_{\mu 2}^\ast
\simeq -2\frac{\Delta E_{21}}{Ac_{13}^2}
c_{13} s_{12} c_{12} c_{23}
\simeq -\frac{\Delta E_{21}}{A}\,c_{23} \sin2\theta_{12}\,,
\label{xem2}
\end{eqnarray}
where $|s_{13}|\ll 1$ was assumed to derive $2\tilde{U}_{e2}\tilde{U}_{\mu 2}^\ast$ in Eq.(\ref{xem2}). Thus we obtain
\begin{eqnarray}
&{\ }&\hspace{-32mm}
P_{\mu e} =
\left|
2\tilde{U}_{e3}\tilde{U}_{\mu 3}^\ast 
\sin\left(\frac{\Delta \tilde E_{31} L}{2}\right)
+e^{i\Delta \tilde E_{32} L/2}
2\tilde{U}_{e2}\tilde{U}_{\mu 2}^\ast 
\sin\left(\frac{\Delta \tilde E_{21} L}{2}\right)
\right|^2
\nonumber\\
&{\ }&\hspace{-25mm}
\simeq
\left|
\frac{\Delta E_{31}}
{\Delta \tilde{E}_{31}^{(0)}}\, e^{-i\dcp} s_{23} \sin2\theta_{13}
\sin\left(\frac{\Delta \tilde E_{31}^{(0)} L}{2}\right)\right.
\nonumber\\
&{\ }&\hspace{-20mm}
\left.+e^{i\Delta \Lambda_+ L/2}
\frac{\Delta E_{21}}{A}\,c_{23} \sin2\theta_{12}
\sin\left(\frac{\Lambda_- L}{2}\right)
\right|^2
\nonumber\\
&{\ }&\hspace{-25mm}
\simeq\left|
\frac{\Delta E_{31}}
{\Delta \tilde{E}_{31}^{(0)}}\, s_{23} \sin2\theta_{13}
\sin\left(\frac{\Delta \tilde E_{31}^{(0)} L}{2}\right)\right.
\nonumber\\
&{\ }&\hspace{-20mm}
\left.+e^{i(\dcp+\Lambda_+ L/2)}
\frac{\Delta E_{21}}{A}\,c_{23} \sin2\theta_{12}
\sin\left(\frac{\Lambda_- L}{2}\right)
\right|^2\,,
\label{pme1}
\end{eqnarray}
where $\Lambda_-$, $\Lambda_+$ and $\Delta \tilde{E}_{31}^{(0)}$ are given by Eqs.\,(\ref{lambdan}), (\ref{lambdap}) and (\ref{deltatildee31}), respectively. This is basically the result obtained by Cervera et al\cite{Cervera:2000kp}. Introducing the notations
\begin{eqnarray}
&{\ }&\hspace{-80mm}
F\equiv \frac{\Delta E_{31}}
{\Delta \tilde{E}_{31}^{(0)}}\,
\sin2\theta_{13}\sin\left(\frac{\Delta \tilde E_{31}^{(0)} L}{2}\right)\,,
\label{f}\\
&{\ }&\hspace{-80mm}
G\equiv
\frac{\Delta E_{21}}{A}\, \sin2\theta_{12}
\sin\left(\frac{\Lambda_- L}{2}\right)\,,
\label{g}
\end{eqnarray}
Eq.\,(\ref{pme1}) can be rewritten as
\begin{eqnarray}
&{\ }&\hspace{-28mm}
P_{\mu e} \simeq
\left|
s_{23}F+e^{i(\dcp+\Lambda_+ L/2)}
\, c_{23} G\right|^2
\nonumber\\
&{\ }&\hspace{-20mm}
=\frac{F^2+G^2}{2}-\frac{F^2-G^2}{2}
\cos 2\theta_{23}
+FG\cos\left(\dcp+\frac{\Lambda_+ L}{2}
\right)\,\sin 2\theta_{23}
\nonumber\\
&{\ }&\hspace{-20mm}
=\frac{F^2+G^2}{2}-
\sqrt{\left(\frac{F^2-G^2}{2}\right)^2
+\left\{FG\cos\left(\dcp+\frac{\Lambda_+ L}{2}
\right)\right\}^2}
\nonumber\\
&{\ }&\hspace{5mm}
\times
\cos\left[2\theta_{23}+\tan^{-1}
\left\{\frac{2FG\cos\left(\dcp+\Lambda_+ L/2\right)}
{F^2-G^2}\right\}
\right]\,.
\label{oct0}
\end{eqnarray}

\subsection{Octant degeneracy}

Now let us consider the case in which octant degeneracy occurs in the neutrino mode:
\begin{eqnarray}
&{\ }&\hspace{-42mm}
P_{\mu e}\left(\dcp=-\frac{\pi}{2},
\theta_{23}=\frac{\pi}{4}-\epsilon\right)
= P_{\mu e}\left(\dcp'=\pi,
\theta_{23}'=\frac{\pi}{4}+\epsilon'\right)\,
\label{degen1}
\end{eqnarray}
where all other parameters are similar, and $\epsilon=3^\circ=\pi/60\ll 1$. In this case, since $\epsilon$ and $\epsilon'$ are small, from Eq.\,(\ref{oct0}) we have approximately
\begin{eqnarray}
&{\ }&\hspace{-45mm}
P_{\mu e}\left(\dcp=-\frac{\pi}{2},
\theta_{23}=\frac{\pi}{4}-\epsilon\right)
\nonumber\\
&{\ }&\hspace{-50mm}
=\frac{F^2+G^2}{2}-
\sqrt{\left(\frac{F^2-G^2}{2}\right)^2
+\left\{FG\cos\left(-\pi/2+\Lambda_+ L/2\right)\right\}^2}
\nonumber\\
&{\ }&\hspace{-25mm}
\times
\sin\left[2\epsilon-\tan^{-1}
\left\{\frac{2FG\cos\left(-\pi/2+\Lambda_+ L/2\right)}
{F^2-G^2}\right\}
\right]
\nonumber\\
&{\ }&\hspace{-50mm}
=\frac{F^2+G^2}{2}+\left|\frac{F^2-G^2}{2}\right|
\sqrt{1+\left\{\frac{2FG\sin\left(\Lambda_+ L/2\right)}
{F^2-G^2}\right\}^2
}
\nonumber\\
&{\ }&\hspace{-25mm}
\times\sin
\left[\tan^{-1}\left\{\frac{2FG\sin\left(\Lambda_+ L/2\right)}
{F^2-G^2}\right\}
-2\epsilon\right]
\nonumber\\
&{\ }&\hspace{-50mm}
\simeq\frac{F^2+G^2}{2}+\left|\frac{F^2-G^2}{2}\right|
\cdot\left\{\frac{2FG\sin\left(\Lambda_+ L/2\right)}
{F^2-G^2}-2\epsilon
\right\}
\,,
\label{degen2r}\\
&{\ }&%\vspace{10mm}
\nonumber\\
\nonumber\\
&{\ }&\hspace{-45mm}
P_{\mu e}\left(\dcp'=\pi,
\theta_{23}'=\frac{\pi}{4}+\epsilon'\right)
\nonumber\\
&{\ }&\hspace{-50mm}
=\frac{F^2+G^2}{2}+
\sqrt{\left(\frac{F^2-G^2}{2}\right)^2
+\left\{FG\cos\left(\dcp'+\Lambda_+ L/2\right)\right\}^2}
\nonumber\\
&{\ }&\hspace{-25mm}
\times
\sin\left[2\epsilon'+\tan^{-1}
\left\{\frac{2FG\cos\left(\dcp'+\Lambda_+ L/2\right)}
{F^2-G^2}\right\}
\right]
\nonumber\\
&{\ }&\hspace{-50mm}
=\frac{F^2+G^2}{2}-\left|\frac{F^2-G^2}{2}\right|\cdot
\sqrt{1+\left\{\frac{2FG\cos\left(\Lambda_+ L/2\right)}
{F^2-G^2}\right\}^2}
\nonumber\\
&{\ }&\hspace{-25mm}
\times
\sin
\left[\tan^{-1}\left\{\frac{2FG\cos\left(\Lambda_+ L/2\right)}
{F^2-G^2}\right\}
-2\epsilon'\right]
\nonumber\\
&{\ }&\hspace{-50mm}
\simeq\frac{F^2+G^2}{2}-\left|\frac{F^2-G^2}{2}\right|
\left\{\frac{2FG\cos\left(\Lambda_+ L/2\right)}
{F^2-G^2} -2\epsilon'\right\}
\label{degen2w}
\end{eqnarray}
Thus from Eqs.\,(\ref{degen2r}), (\ref{degen2w}) and (\ref{degen2w}) we have
\begin{eqnarray}
&{\ }&\hspace{-45mm}
\epsilon' \simeq
-\epsilon + \frac{FG}{F^2-G^2}
\left\{
\sin\left(\frac{\Lambda_+ L}{2}\right)
+\cos\left(\frac{\Lambda_+ L}{2}\right)
\right\}\,.
\nonumber
\end{eqnarray}
Let us assume the oscillation maximum at T2HK. For $\epsilon=\pi/60 = 3^\circ$, we obtain
\begin{eqnarray}
&{\ }&\hspace{-100mm}
\epsilon'=0.0828 = 4.7^o\,.
\nonumber
\end{eqnarray}
However, for DUNE, again assuming the oscillation maximum, we get
\begin{eqnarray}
&{\ }&\hspace{-100mm}
\epsilon'=0.0453 = 2.6^o\,.
\nonumber
\end{eqnarray}
These analytical discussions for octant degeneracy do not quantitatively agree with the numerical result; however, they are useful for helping us to understand that the fake value $\epsilon'$ does not vanish even in the limit $\epsilon\to 0$.

%\clearpage
%=======================
\bibliographystyle{JHEP}
\bibliography{T2HK_DUNE_density}
%=======================

%============
\end{document}